\documentclass[10pt,%
superscriptaddress,
amsmath,amssymb,amsfonts,
twocolumn,
aps,
]{revtex4-2}

\usepackage{amssymb}
\usepackage{amsfonts}
\usepackage{graphicx}  
\usepackage{amsmath,amssymb}
\usepackage{physics}          
\usepackage{amsthm}
\usepackage{natbib}
\usepackage{empheq}
\usepackage[x11names, svgnames, dvipsnames]{xcolor}
\usepackage{mathtools}
\usepackage{framed}
\usepackage{animate}
\newcommand{\eps}{\varepsilon}

\newcommand{\G}{\mathbb{G}}
\newcommand{\F}{\mathbb{F}}

\newcommand{\Proj}{\mathbb{P}}

\newcommand{\I}{\mathbb{I}}
\newcommand{\Id}{\mathbb{I}}

\newcommand{\Sel}{\mathbb{S}}

\newcommand{\EX}{\mathbb{E}}

\newcommand{\ve}{\vb{e}}

\newcommand{\mI}{\vb{I}}

\newcommand{\vj}{\vb{j}}

\newcommand{\vn}{\vb{n}}
\newcommand{\Gt}{\G_{t,RS}}

\newcommand{\Ft}{\F_{t,RS}}

\DeclareMathOperator{\sinc}{sinc}
\DeclareMathOperator{\diagop}{diag}
\DeclareMathOperator{\vecop}{vec}

\DeclarePairedDelimiter{\avg}{\langle}{\rangle}
\DeclarePairedDelimiterX{\inner}[2]{\langle}{\rangle}{#1,#2}
 \usepackage[normalem]{ulem}
\usepackage{cancel}

\theoremstyle{plain}

\theoremstyle{definition}

\theoremstyle{remark}

\begin{document}
\title{Optically Incoherent Photonic Mutual Information}

\author{Francis~J.~Chen}
\email{francis.chen@princeton.edu}
\affiliation{Department of Electrical and Computer Engineering, Princeton University, Princeton, New Jersey 08544, USA}

\author{Alessio~Amaolo}
\affiliation{Department of Chemistry, Princeton University, Princeton, New Jersey 08544, USA}

\author{Pengning~Chao}
\altaffiliation{Present address: Maxwell Labs, St. Paul, MN, USA}
\affiliation{Department of Mathematics, Massachusetts Institute of Technology, Cambridge, Massachusetts 02139, USA}

\author{Sean~Molesky}
\affiliation{Department of Engineering Physics, Polytechnique Montréal, Montréal, Québec H3T 1J4, Canada}

\author{Zin~Lin}
\affiliation{Bradley Department of Electrical and Computer Engineering, Virginia Tech, Blacksburg, VA 24060, USA}

\author{Alejandro~W.~Rodriguez}
\affiliation{Department of Electrical and Computer Engineering, Princeton University, Princeton, New Jersey 08544, USA}

\date{\today}
\begin{abstract}
While traditional evaluations of optical information transfer rely on disjointed abstractions to bridge electromagnetic propagation, coherence, and communication theory, we introduce an end-to-end framework that directly connects rigorous subwavelength wave physics to Shannon mutual information. By lifting the Maxwell current-to-field Green's function to propagate second-order field correlations (the mutual intensity), we establish a unified linear channel model that encapsulates coherent communication, phase retrieval, and incoherent imaging. Source-coherence restrictions and receiver-readout projections select the relevant map from this construction, while the full noisy square-law observation law remains nonlinear. Applying this framework, we demonstrate that the mutual-information-optimized photonic front end is dictated jointly by available spatial degrees of freedom, source statistics, and detection laws. For coherent sources measured by square-law detectors, we identify a structural transition: when detectors outnumber sources, topology-optimized front ends shift from point-focusing to interferometric mixing. This mixing leverages interference cross terms to make relative source phases information-bearing, yielding mutual information that surpasses the point-focusing amplitude-only baseline. Conversely, for spatially incoherent sources, the channel collapses to the Hadamard square of the Green's function. In this regime, for equal source and detector counts under an isotropic source covariance, we prove that point-focusing uniquely maximizes the mutual information at fixed Frobenius norm. Under source correlations, the optimized front ends depart from focusing toward distributed optical mixing. Finally, we derive closed-form upper bounds on incoherent mutual information that apply to any photonic front end, governed entirely by the coherent singular values of the underlying electromagnetic operator. Potential applications include near-field microscopy, direct-detection optical datalinks, reference-free phase retrieval, fluorescence and thermal imaging, and structure-agnostic benchmarks for end-to-end-designed computational imagers.
\end{abstract}

\maketitle

\section{Introduction} \label{sec:intro}

Should an optical front end route each point of a scene to its own detector, or should it mix light from many points before recording intensity? Traditional imaging answers ``route'': a lens forms an image by mapping object points to detector pixels, so the recoverable information is ultimately limited by how finely the optics separates neighboring points. A growing body of work instead answers ``mix.'' Lensless cameras, coded apertures, diffractive and metasurface optics, and end-to-end learned optical front ends deliberately redistribute the optical field across the sensor and recover the scene computationally~\cite{gottesman1989new,asif2017flatcam,antipa2018diffusercam,sitzmann2018end,lin2021end,lin2022end,fenimore1978coded}. Despite this proliferation of engineered, inverse-designed, and learned optical architectures, a basic question remains unresolved: \textit{When is optical mixing information-theoretically optimal, when is it merely acceptable, and when is focusing not simply convenient but the best that any optical system can do?}

Behind it lies the deeper, fundamental question: \textit{What are the achievable limits of electromagnetic information transfer?} Imaging, measurement, and communication all broadly seek to approach these limits within their respective operating regimes. However, definitions of resolution spot size, spatial multiplexed throughput, bandwidth, and signal-to-noise ratio exist as field-specific, heuristic criteria that do not individually define the information gain an optical system can transfer from a sender or object, to a receiver or detector. To that end we propose an end-to-end framework that bridges these disparate figures of merit across different subfields through information-theoretic analysis, while modeling the full extent of wave propagation under Maxwell's equations that govern them. This model includes the considerations of optical coherence, wave scattering physics, geometric limits, and correlations within the ensembles formed by the objects and senders.

Communication theory measures what any channel can convey by the mutual information \(\mathcal I(\vb{x};\vb y)\) between an input \(\vb{x}\) and a measurement \(\vb{y}\): the average information over possible inputs that the measurement reveals about each input. The mutual information is fixed by the input probability distribution together with the channel law, the conditional distribution of \(\vb{y}\) given \(\vb{x}\) that specifies the statistics of the measurement outcomes~\cite{cover2006elements}. The connections between imaging, measurement, and communication are as old as information theory~\cite{shannon1948mathematical}: Hartley's founding paper treated picture transmission and television alongside telegraphy, with image blur as a form of intersymbol interference~\cite{hartley1928transmission}, and the assessment of optical images by their information content followed~\cite{fellgett1955assessment,toraldo1955resolving,gabor1961light}. Moreover, mutual information directly depends on spatial resolution: given a finite field-of-view, the compactness of an imaging system's point-spread-function (PSF) defines the per-image throughput of any optical system, similar to the field modes available to a multiplexing antenna~\cite{poon2005degrees,migliore2006role}.

Since the input and the measurements are physical electromagnetic objects, scattering and propagation under Maxwell's equations restrict this throughput on both sides: the realizable input statistics under optical coherence, and the number of field patterns the measurements can distinguish. A planar aperture supports at most as many orthogonal field modes as its space--bandwidth product permits~\cite{lohmann1996space,gustafsson2025shadow,gustafsson2025dofradiating}, and the angular spectrum of the field---its decomposition into plane waves~\cite{goodman2017introduction}---divides those modes into radiative components, which propagate to large separations, and evanescent components, which carry subwavelength detail but decay below the noise floor within roughly a wavelength. Here the communication and sensing communities diverge: considerations of wavelength in relation to the geometry of each community's typical operating regimes, as well as the physical variables connected by the coherent channel matrix or incoherent PSF, lead to differing criteria of information transfer.  

Classical antenna theory treats the underlying wave physics as a fixed channel matrix from the far-field propagation of radiating modes in vacuum or uncontrolled environments, decomposing a field channel into singular sub-channels above the noise floor and allocating power among them~\cite{shannon1949communication,  foschini1998limits,telatar1999capacity}. Fourier optics and wide-field microscopy similarly discard the use of evanescent modes and the subwavelength-scale structuring in engineering the PSF~\cite{abbe1873beitrage,shechtman2014optimal}, relying on manipulation of propagating wavevectors for super-resolution~\cite{gustafsson2000surpassing}, and expressing limitations through numerical aperture, Abbe or Rayleigh resolution, focal distance, point-spread-function width, and aperture to arrive at the space--bandwidth product~\cite{abbe1873beitrage,rayleigh1879investigations,fellgett1955assessment,born1999principles,goodman2017introduction}. Compressed sensing and computational imaging emphasize the conditioning of a measurement matrix together with the structure of the source prior (i.e. the assumed distribution of the object ensemble)~\cite{antipa2018diffusercam,candes2006near,duarte2011structured,kabuli2026designing,donoho2006compressed}. 
Phase retrieval relates intensity measurements to real- or complex-valued unknowns modulo sign or phase ambiguities~\cite{shechtman2015phase,bandeira2014saving}. Each criterion is appropriate within its own regime of Maxwell's equations (ray optics~\cite{born1999principles}, far-field propagation~\cite{goodman2017introduction}, quasistatics~\cite{novotny2012principles,chao2022physical,gustafsson2016physical}) but rests on abstractions valid only within its scope; the assumptions differ in the source variables, the accessible field correlations, and the measured detector variables.

Throughout this paper, we introduce a description of the measurable electromagnetic degrees of freedom selected jointly by source coherence, object-ensemble correlations, and the detector law. We take the mutual information \(\mathcal{I}(\vb{x};\vb{y})\) as the figure of merit for every channel law we introduce, bridging the criteria and physical abstractions across various modes of communication and sensing.
Each channel law is united by one underlying physical map: propagation under Maxwell's equations carries information about a sender or object, represented by equivalent source currents, to fields at the detectors, not only through the physical quantities themselves, but also via their mutual phase relationships. The detection mechanism then selects which projection of that map becomes the information channel. Coherent communication, phase retrieval, and incoherent intensity imaging are therefore different source and receiver projections of the same Maxwell current-to-field operator and its lifted coherence operator (see Sec.~\ref{sec:channel_laws}). 

Figure~\ref{fig:phasespace} displays the hierarchy of channel laws obtained from a single electromagnetic Green's function.  For a specified structure, the Green's function gives the field produced at a receiver point by a point current source, and \(\Gt\) denotes the corresponding source-to-receiver block. Our prior full-wave Shannon-capacity framework~\cite{amaolo2026maximum} treated the phase-sensitive, coherent limit of this hierarchy: the source controls complex current amplitudes, the receiver measures complex fields, connected by a linear channel law. Within the coherent limit, nanophotonic structuring shapes the channel matrix, and, when optimized alongside the signal covariance, answers the question: \textit{to what extent can photonic structuring maximize the mutual information of a coherent electromagnetic system subject to size, material, and bandwidth constraints?} In parallel, recent works have employed end-to-end inverse design~\cite{lin2021end,lin2022end,pestourie2023efficient}, rank-relaxation bounds~\cite{molesky2022t,chao2022physical,kuang2025bounds}, and information-driven measurement design~\cite{markley2024information} to quantify how structuring may enhance reconstruction, beginning to address a related question: \textit{how many information channels can a photonic structure support?} 

The present work asks the complementary question of how information can be encoded electromagnetically beyond the coherent fields themselves, namely, in the second-order statistics. These statistics enter on two timescales: optical coherence sets the correlations among the fields within a single channel use, and the source distribution sets their variation across uses. This model changes how electromagnetic propagation can be decomposed into information channels, and in specific limits the resulting operators admit the same spectral decomposition as the coherent Green's function for bounding the mutual information. For instance, fully incoherent sources restrict the encoding of information to the diagonal elements of the coherence matrix, while square-law detection accesses only the diagonal elements of the field coherence matrix at the detector, giving rise to the linear intensity operator \(\Ft=\Gt^*\odot\Gt\), where \(\odot\) is the entrywise (Hadamard) matrix product. Inequalities for Hadamard products bound the singular values of \(\Ft\) by those of \(\Gt\)~\cite{ando1987singular,horn1991topics}. More generally, however, mutual coherence and detected intensities are quadratic in the field, and the electromagnetic mutual information is subsequently shaped beyond the capacity-achieving distributions of well-behaved Gaussians. Yet square-law detection eliminates Gaussian statistics: even Gaussian-distributed fields produce non-Gaussian-distributed intensities, and the mutual information has no closed form~\cite{shlezinger2018measurement}. Our information-theoretic extension of the full physics of electromagnetism to incorporate currents, fields, and intensities of varying correlations and coherence allows for the rigorous treatment of problems beyond phase-sensitive communication, encompassing wide-field imaging, microscopy, compressed sensing, phase retrieval, phase-space tomography, and direct-detection datalinks.

Our framework has two consequences. First, it places on one footing a growing body of work joining information theory to electromagnetism---the wave theory of information~\cite{franceschetti2017wave}, communication-modes analyses~\cite{miller2019waves}, end-to-end and computational imaging~\cite{sitzmann2018end,antipa2018diffusercam,markley2024information,kabuli2026designing}, and full-wave Shannon capacity over structured media~\cite{amaolo2026maximum}---across regimes of coherence and propagation. Second, because the channel operators are sub-blocks and selections of one Green's function, photonic inverse design~\cite{molesky2018inverse}, standard information-theoretic machinery~\cite{cover2006elements,telatar1999capacity}, and the results of the coherent-capacity framework~\cite{amaolo2026maximum} act on every projection. In the two intensity-detection limits treated here, these methods already yield exact high-SNR laws, a numerical focusing-to-interferometric transition for phase retrieval as detectors are added, the conditional optimality of point focusing for incoherent imaging, and closed-form mutual information upper bounds on any incoherent meta-imager.

We apply our formulation to address several problems in imaging, sensing, and communication. Under square-law detection, the next two subsections group these problems into two regimes of coherence: coherent sources, covering phase retrieval and direct-detection optical communication, and spatially incoherent sources, covering fluorescence, thermal, and computational imaging, as well as wireless links in the radiating near field. The derivations appear in Secs.~\ref{sec:phase_retrieval} and~\ref{sec:incoherent_MI}.

\subsection{Phase Retrieval and Direct-Detection Communication}

Classical phase retrieval concerns the measurement of coherent phase relationships using intensity detection \cite{shechtman2015phase}, without the aid of a reference beam or local oscillator. Existing theoretical works determine the minimum number of intensity measurements required to determine the source field, and with it the coherence matrix, up to a global phase~\cite{balan2006signal,balan2015invertibility,conca2015algebraic,bandeira2014saving,vinzant2015small}. The abstract measurement matrix that connects coherent sources with intensity readouts has been optimized under trace norm budgets~\cite{shlezinger2018measurement}, but without consideration of constraints introduced by physical propagation from the sources through structured optics. 

For point coherent sources measured by an array of point intensity detectors of varying count, we topology-optimize photonic front ends for mutual information from first principles, validating well-known mathematical models of classical phase retrieval~\cite{shechtman2015phase,shlezinger2018measurement}: the optima split between focusing and interferometric mixing, depending on the ratio of intensity detectors to coherent sources. When the number of physical detectors \(M_R\) exceeds the number of distinct sources \(M_S\), the mutual-information-optimized front end becomes interferometric. We demonstrate this transition numerically for \(M_S=2\), optimizing photonic front ends for mutual information at each detector count \(M_R\) from \(1\) to \(8\), arriving at topology-optimized structures that point focus at \(M_R=M_S\) and are interferometric for \(M_R>M_S\), exceeding the mutual information of the amplitude-only point-focusing configuration. When the intensity detectors do not outnumber the coherent sources (\(M_S\geq M_R\)), we derive the high-SNR mutual information of the mode-sorted channel (each independent field mode routed to its own detector) up to a vanishing remainder for complex Gaussian inputs (Eq.~\eqref{eq:half_logdet_asymptotic_diagonal_hiSNR}). At \(M_R=M_S=1\) the law reduces to the classical half pre-log---the coefficient of \(\log\mathrm{SNR}\) in the high-SNR mutual information~\cite{durisi2011high}---of an amplitude modulation (direct-detection) receiver connected with a sender modulating both its amplitude and phase~\cite{blachman1953comparison}. While mutual information of phase retrieval settings (see Eq.~\eqref{eq:ICAWGN_general_mutual_info_bound_objective}) has no closed-form expression and is generally computationally costly to optimize~\cite{shlezinger2018measurement}, we construct a Gaussian-matched surrogate program subject to scattering constraints and heteroscedastic (signal-dependent-variance) noise used only to supply gradients, and report the mutual information value corresponding to the optimized structure as a Monte Carlo estimate of the exact integral (Appendix~\ref{asec:surrogate_objective}).

\subsection{Incoherent Imaging and MIMO}

Incoherent imaging and wireless multiple-input multiple-output (MIMO) theory alike hold the propagation fixed: senders and receivers are prescribed, the environment is given, and the channel matrix follows from the arrangement. Beyond the vacuum far field, near fields, evanescent components, and photonic resonances carry information that conventional far-field, shift-invariant point-spread-function models omit~\cite{novotny2012principles,ji2023extra,virally2026many}, and the front end structure sets the channel matrix rather than a fixed convolution kernel. These are the conditions of current practice: lensless cameras record in the Fresnel region of a structured mask~\cite{asif2017flatcam,antipa2018diffusercam}, wireless links with electromagnetically large arrays extend into the radiating near field~\cite{cui2023near,zhang2023focusing}, and near-field microscopy~\cite{betzig1992near}, integrated detectors and multiplexers~\cite{ahn2010evanescent}, and on-chip reconstruction couple directly to evanescent components.

Rather than taking the transfer matrix and PSF as given~\cite{li2020capacity,hranilovic2006pixelated,goodman2017introduction}, we derive a channel law and PSF determined by statistical optics and wave propagation accounting for the full evanescent spectrum in the near field. Our unified Maxwell propagator channel model for coherence matrices at the source and detector reduces to a map across their respective diagonal elements under a temporal average enforcing source incoherence as well as intensity detection. We therefore analytically derive the incoherent PSF operator $\Ft$ from the lifted Maxwell operator (Sec.~\ref{sec:theory}). For a spatially incoherent source with isotropic statistics, measured by as many intensity detectors as sources, we analytically show that point focusing uniquely maximizes the mutual information at every SNR whenever structuring can flatten the singular-value spectrum of the intensity operator while preserving its Frobenius norm (Sec.~\ref{sec:incoherent_MI}). In the radiative far field, the numerical-aperture resolution limit of conventional imaging is classically a statement about point-spread-function overlap~\cite{abbe1873beitrage,rayleigh1879investigations}; we verify the same limit as the information optimum for sources on the diffraction-limited lattice, arriving at the point-focusing configuration that maximizes the mutual information over passive detector-side front ends (Appendix~\ref{asec:farfield_focusing}).

For optically incoherent datalinks, we construct a Gaussian model for the intensity fluctuations about a positive mean: the operator \(\F_{t,RS}\) is necessarily entrywise nonnegative, but the mutual information is invariant to the mean shift, so negative-valued fluctuations are admissible within physical limits, and the linear Gaussian channel model holds whenever negligible probability mass lies outside the nonnegative support (Sec.~\ref{sec:incoherent_MI}). From the same channel we derive a closed-form upper bound: for every source covariance and every noise level, the mutual information of an incoherent front end is bounded through the coherent singular values of its Green's function (Eq.~\eqref{eq:F_MI_ceiling}). The structure-agnostic electromagnetic limits of Refs.~\cite{amaolo2026maximum,virally2026many}, derived for the coherent operator alone, transfer to intensity detection: we bound the singular values of the Hadamard operator by those of the Green's function (Sec.~\ref{sec:incoherent_MI}), converting those limits into mutual information upper bounds on any incoherent meta-imager. 

Under correlated source statistics, the point-focusing optimality guarantee of the isotropic case is lost. At fixed power per mode, the mutual information is largest when the source covariance is diagonal in the right-singular basis of \(\F_{t,RS}\), at every SNR~\cite{telatar1999capacity,carson2012communications}; an isotropic covariance matrix satisfies this alignment in every basis. For a fixed and correlated covariance, however, the alignment must come from the channel side. We demonstrate the departure numerically with topology-optimized structures: under correlated source covariances the optimized front end distributes each source across several detectors rather than point focusing (Fig.~\ref{fig:iso_vs_corr}).

~\\~\\ \paragraph{Concurrent Work.} While this work was in preparation, Ref.~\cite{kienesberger2026end} showed independently that data-free Shannon and Fisher objectives for a nonnegative intensity-transfer matrix are maximized by focusing, routing fields from each source to be concentrated onto a distinct detector. 
Our incoherent-imaging channel reaches a compatible conclusion under isotropic source statistics (Sec.~\ref{sec:incoherent_MI}), here as one regime of the coherence-selected construction: the intensity-transfer matrix is the Maxwell-constrained Hadamard (element-wise) square of the Green's function $\F=\G^{*}\odot\G$ rather than a nonnegative matrix subject to an imposed passivity (column-sum) budget; focusing remains the isotropic-incoherent optimum under the spectrum-flattening condition at fixed Frobenius norm established below, while correlated source priors, and phase-bearing inputs as in the phase-retrieval channel of Sec.~\ref{sec:phase_retrieval}, move the optimum away from focusing. The two results are complementary: focusing is selected by amplitude-only source variables with uncorrelated intensities, interferometry by phase-bearing ones.

\section{Theory and Definitions} \label{sec:theory}
\subsection{Information-Theoretic Channel Model.}
A general MIMO channel model between \(M_S\) source signals of interest and \(M_R\) measured detector signals is defined as
\begin{equation} \label{eq:channel_law_universal}
    \vb{y} = f(\vb{H}\vb{x}+\vb{n}), \;\, \vb{x} \in \mathbb{K}^{M_S}, \; \vb{n} \in \mathbb{K}^{M_R}, \; \vb{y} \in f(\mathbb{K})^{M_R},
\end{equation}
where \(f\) denotes an element-wise function potentially capturing nonlinear aspects of the channel, \(\vb{x} \sim \gamma_{\vb x}, \vb y \sim \gamma_{\vb y}, \vb n \sim \mathcal{KN}(0,N \I)\) for some total noise \(N\) are random variables that represent the input signal, the output signal, and additive white Gaussian noise of variance \(N\) (i.e., AWGN with $\mathbb{E}[n_i n_j^{*}] = N\delta_{ij}$, with total variance
$\mathbb{E}[|n_i|^{2}] = N$ per component), respectively, and \(\mathbb{K}\) stands for \(\mathbb{C}\) or $\mathbb{R}$. Setting \(f\) to the identity operation and \(\mathbb K = \mathbb C\) for all signals recovers a well-known communication framework with a simple linear channel model~\cite{telatar1999capacity,tse2005fundamentals,poon2005degrees}, which we below define to be a \textit{coherent-to-coherent} communication framework.

In communication, sensing, or imaging, a key figure of merit is the mutual information \(\mathcal I\), which dictates the information gained about \(\vb x\) by measuring \(\vb y\).
Encoding of the input statistics shapes the marginal densities \(\gamma_{\vb{x}}\).
The Shannon capacity is the supremum of communication rates achievable with arbitrarily small error probability and equals \(\sup_{\gamma_{\vb{x}}} \mathcal I\) under a power constraint on \(\vb{x}\).
\(f\), noise statistics, and the design of the physical channel \(\vb{H}\) collectively shape the conditional densities \(\gamma_{\vb{y}|\vb{x}}\). 
Together, these quantities determine the mutual information
\begin{equation} \label{eq:mutual_info_integral_universal}
    \mathcal{I}(\mathbf{x}; \mathbf{y}) = \int_{\Omega} \gamma_{\mathbf{x}}(\mathbf{x}) \, \gamma_{\mathbf{y}|\mathbf{x}}(\mathbf{y}|\mathbf{x}) \log \left[ \frac{\gamma_{\mathbf{y}|\mathbf{x}}(\mathbf{y}|\mathbf{x})}{\gamma_{\mathbf{y}}(\mathbf{y})} \right] \dd \mathbf{y} \dd\mathbf{x},
\end{equation}
where \(\Omega=\operatorname{supp}(\gamma_{\vb x})\times\operatorname{supp}(\gamma_{\vb y})\) (we use the natural logarithm throughout, so mutual information is measured in nats).
In general, this integral is intractable, but for $f$ the identity and a zero-mean complex Gaussian-distributed input with covariance \(\vb Q := \mathbb{E}[\vb x \vb x^{\dagger}]\), the densities \(\gamma_{\vb{x}}, \gamma_{\vb{y}}, \gamma_{\vb{y}|\vb{x}}\) are all Gaussian, and under the average-power constraint \(\Tr[\vb Q]\leq P\) the maximum reduces to
\begin{equation} \begin{aligned} \label{eq:logdet}
    C &= \sup_{\gamma_{\vb x}} \mathcal I \xrightarrow[{\Tr[\vb{Q}] \leq P}, \, \mathbb K=\mathbb C]{\gamma_{\vb x} \sim \mathcal{CN}, \gamma_{\vb y} \sim \mathcal{CN}, \gamma_{\vb{y}|\vb{x}}=\gamma_{\vb{n}}} \\
    & \quad \max_{\vb Q\succeq 0,\,\Tr[\vb Q]\leq P} \log\det \left( \I + \frac{1}{N}\vb{H}\vb{Q}\vb{H}^{\dagger} \right), 
\end{aligned} \end{equation}
a well-known result in information theory for coherent-to-coherent communication and the basis for a variety of information-theoretic results~\cite{telatar1999capacity}.
For fixed channel matrices \(\vb H\), the supremum over \(\vb Q\) is a convex problem that can be solved with a procedure known as waterfilling~\cite{palomar2003joint}.
While analytically tractable, the assumption of both input and output random variables distributed normally is not always valid for all channel laws. For instance, when \(f\) enacts element-wise square-law detection, a Gaussian density \(\gamma_{\vb{x}}\) and additive pre-detection Gaussian noise \(\gamma_{\vb{n}}\) induce a noncentral chi-squared distribution for \(\gamma_{\vb{y}|\vb{x}}\)~\cite{goodman2015statistical}, leaving \(\mathcal{I}(\vb{x};\vb{y})\) expressible only as an integral that must be evaluated numerically~\cite{shlezinger2018measurement}. To address this gap, we seek to develop a theory of incoherence in information-theoretic terms.

\subsection{Electromagnetic Communication Channels}
The abstract channel matrix \(\vb H\) in Eq.~\eqref{eq:channel_law_universal} is constrained in a photonic system by Maxwell's equations~\cite{gustafsson2011physical,ehrenborg2018fundamental,ehrenborg2020radiation,ehrenborg2021capacity,ivrlac2010toward}. 
Let \(S\) denote the sender, object, or source region and \(R\) the receiver or detector region. 
At a single angular frequency \(\omega\), an information-bearing input field is represented physically by a free-current vector \(\vj_i\) supported in \(S\), and the corresponding pre-detection output is the electric field \(\ve_R\) incident on \(R\). 
We write \(Z=\sqrt{\mu_0/\varepsilon_0}\) for the vacuum impedance and \(k=\omega/c\) for the vacuum wavenumber; in the dimensionless units used throughout, \(\mu_0=\varepsilon_0=1\), so \(Z=1\), \(c=1\), and \(k=\omega\). 
The frequency-domain electric field obeys
\begin{equation}
\label{eq:maxwell_current_source}
\left[\nabla\times\nabla\times-\eps(\vb r)\omega^2\right]\ve(\vb r;\omega)
= i\omega\,\vj_i(\vb r),
\end{equation}
where \(\eps(\vb r)\) is the relative permittivity profile. 
The dyadic Green's function \(\G_t(\vb r,\vb r';\eps;\omega)\) is the response kernel of the structured environment, defined by
\begin{equation}
\label{eq:green_function_proper}
\left[\nabla\times\nabla\times-\eps(\vb r)\omega^2\right]
\G_t(\vb r,\vb r';\eps;\omega)
=\omega^2\mI\,\delta(\vb r-\vb r'),
\end{equation}
so that the field generated by an arbitrary source current is
\begin{equation}
\label{eq:green_convolution}
\ve(\vb r;\omega)
=
\frac{i}{\omega}\int_S 
\G_t(\vb r,\vb r';\eps;\omega)\,\vj_i(\vb r')\,\dd\vb r' .
\end{equation}
The subscript \(t\) emphasizes that this is the total-field Green's function, including the incident field and all re-radiation from the structured medium. 
In vacuum, \(\eps=\Id\) and we write \(\G_{\mathrm{vac}}\equiv \G_t(\vb r,\vb r';\Id;\omega)\). 
When the structure is parameterized by a susceptibility \(\chi_D\) in a design region \(D\), we write
\begin{equation}
\eps(\vb r)-\Id=\Proj_{\eps}(\vb r)\chi_D,
\end{equation}
with \(\Proj_{\eps}\) carrying the spatial support and layout of the design.

\subsection{Coherence-based Channel Law} \label{sec:channel_laws}

Sampling Eq.~\eqref{eq:green_convolution} on \(M_S\) source degrees of freedom in \(S\) and \(M_R\) receiver degrees of freedom in \(R\) yields the bare, \(\omega^2\mI\delta\)-normalized Green's-function matrix
\(\tilde{\G}_{t,RS}\in\mathbb C^{M_R\times M_S}\). 
The finite-dimensional current-to-field channel used in the information-theoretic model absorbs the propagation prefactor, i.e., \(\Gt := \frac{i}{\omega}\,\tilde{\G}_{t,RS}\), \(\ve_R=\Gt\,\vj_i \).
Thus, before detector readout, Eq.~\eqref{eq:channel_law_universal} specializes to
\[
    \vb x\equiv \vj_i,\qquad
    \vb H\equiv \Gt,\qquad
    \vb H\vb x\equiv \ve_R .
\]
For notational economy we abbreviate \(\Gt\) by \(\G\) when the source and receiver regions are clear. 
The distinction between \(\tilde{\G}_{t,RS}\) and \(\Gt\) is only the fixed scalar \(i/\omega\): \(\tilde{\G}_{t,RS}\) is the sampled Green's function proper, whereas \(\Gt\) is the physical channel that maps current amplitudes directly to receiver fields. 
We keep \(\tilde{\G}_{t,RS}\) only when invoking singular-value bounds on the Green's function itself~\cite{amaolo2026maximum}; the phase-retrieval and incoherent channels below are written through the single current-to-field channel \(\Gt\).

The remaining differences among the channel laws do not alter Maxwell propagation, instead altering which source statistics and receiver variables are available, as the same current-to-field map also propagates the mutual intensities of the fields. Writing $\avg{\cdot}$ for the temporal average over the detector integration time spanning one use of the channel, distinct from the ensemble expectation $\EX[\cdot]$ over the signal distribution~\cite{goodman2015statistical}, the source and receiver coherence matrices obey
\[
\avg{\ve_R\ve_R^\dagger}=\Gt\,\avg{\vj_i\vj_i^\dagger}\,\Gt^\dagger,
\]
or, in row-major vectorization, with \(\vecop\) listing the entries of a matrix row by row in a single column,
\[
\vecop\!\left(\avg{\ve_R\ve_R^\dagger}\right)=\left(\Gt\otimes\Gt^*\right)\vecop\!\left(\avg{\vj_i\vj_i^\dagger}\right).
\]
The dimension-lifted Kronecker product $\Gt\otimes\Gt^*$ describes how source coherence is converted into receiver coherence, while the detector law selects which components are recorded.

A phase-sensitive receiver keeps the complex field \(\ve_R+\vn\), so \(f\) is the identity. 
A square-law receiver applies \(f(\cdot)=|\cdot|^{\circ2}\), producing the phase-retrieval channel of Sec.~\ref{sec:phase_retrieval}. 
A spatially incoherent source imposes the temporal-average restriction
\(\avg{\vj_i\vj_i^{\dagger}}=\diagop(\vb B)\), and, together with square-law readout, collapses the lifted coherence propagation to the Hadamard point-spread-function channel
\(\Ft=\Gt^{*}\odot\Gt\), used in Sec.~\ref{sec:incoherent_MI}. Coherent communication, phase retrieval, and incoherent intensity imaging thus share the same underlying field propagation, the current-to-field map \(\Gt\) and its mutual intensity lift \(\Gt\otimes\Gt^{*}\), and differ only in the source statistics admitted and the receiver variables retained.

\subsection{Informational Degrees of Freedom} \label{sec:MRprimeMSprime}
Because these projections change the physical input and output variables on both sides of the channel law, the electromagnetic dimensions relevant to information transfer are not \(M_S\) and \(M_R\) themselves, but the effective source and measurement dimensions induced by the selected channel law.
We denote the source complexity by \(M_S'\), the number of real source parameters per channel use that the ensemble varies and that the selected detector law can distinguish. 
When the source model is described by a covariance, we write this as \( M_S'\approx \operatorname{rank}_{\mathbb R}(\vb Q_x)\), evaluated over the directions the ensemble varies and the detector law can distinguish.
Source correlations reduce this dimension, spatial incoherence removes the source phases from the varied parameters, and square-law detection removes source transformations that leave every noiseless intensity measurement unchanged, such as a global phase.

Similarly, \(M_R'\) denotes the effective real dimension of the measurement variables that the apparatus can resolve under the selected channel law: \(M_R'\approx \operatorname{rank}_{\mathbb{R}}(\vb H),\)
where \(\vb H\) is the operator acting on the selected source and measurement variables: \(\Gt\) for coherent field readout, a receiver-diagonal projection of \(\Gt\otimes\Gt^*\) for coherent square-law readout, and \(\Ft\) for spatially incoherent intensity readout.  This rank is only approximate, as considerations for correlations and a large condition number decrease the effective rank: two maximally correlated sources share a single degree of freedom, while a sub-channel of the channel operator---one input--output mode pair of its singular-value decomposition---delivers no information once its singular value falls below the noise floor. Thus \(M_S'\) and \(M_R'\) are not intrinsic matrix ranks; they are model-dependent effective dimensions that summarize the source statistics, Maxwell propagation, detector law, and noise level in a common phase-space representation.

\begin{figure*}[t]
  \centering
  \includegraphics[width=\textwidth]{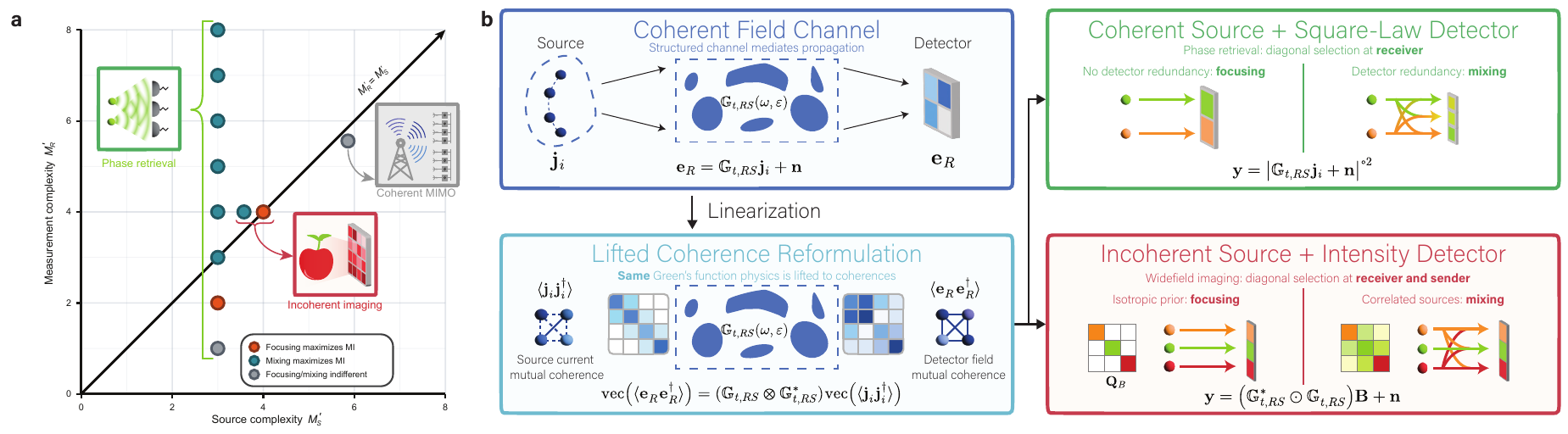}
  \caption{\textbf{``Imaging Complexity'' phase space.} (a) Systems are placed by the effective real source dimension ($M_S'$) and the effective real measurement dimension ($M_R'$), both per channel use. Marker color denotes the mutual-information optimum from results obtained in this work. Coherent MIMO is indifferent (gray): linear phase-sensitive detection depends on the Green's function only through its singular values, and the entrywise arrangement separating focusing from mixing is invisible to it. The phase-retrieval column ($M_S'=2M_S-1=3$) is indifferent at the single scalar measurement $M_R=1$, focusing at $M_R=2$, and mixing from the diagonal onward; incoherent imaging at $M_R=M_S=4$ is focusing under an isotropic prior and mixing under a correlated one, whose poor conditioning also lowers $M_S'$. 
  (b) One Green's function generates every channel law shown: $\Gt$ propagates source currents to receiver fields (coherent channel), and its lift $\Gt\otimes\Gt^{*}$ propagates the source coherence matrix to the receiver coherence matrix. Square-law detection of a coherent source selects the receiver diagonal ($\Sel_R$, Sec.~\ref{sec:channel_laws}), the phase-retrieval channel; source incoherence selects both diagonals ($\Sel_R$, $\Sel_S$), the Hadamard channel acting on the source intensities ($\vb B$).}
  \label{fig:phasespace}
\end{figure*}

Fig.~\ref{fig:phasespace} places imaging, sensing, and communication systems on a common source--measurement plane, with these effective dimensions as coordinates. The diagonal \(M_R'=M_S'\) marks a parameter-matching boundary: below it the source ensemble outnumbers the measurements, so aggregation, prior information, or multiple channel uses become necessary. The boundary is quantitative: at high SNR each real degree of freedom conveys at most one half of \(\log\mathrm{SNR}\), so the mutual information grows at most as \(\tfrac{1}{2}\min(M_S',M_R')\log\mathrm{SNR}\)~\cite{foschini1998limits,telatar1999capacity,tse2005fundamentals,lozano2005high}. For a wide-field incoherent microscope, \(M_R'\) is the photodetector count in the aperture and field of view, reduced to the number of usable modes of \(\Ft\), and \(M_S'\) is the number of statistically independent source-intensity variables in the object ensemble, not the number of grid points used to discretize the object.

~\\ \paragraph{Electromagnetic Sensing and Communication.}
The mutual information of every channel below is evaluated under an input distribution \(\gamma_{\vb{x}}\) known to the receiver. In imaging and sensing the source statistics are given rather than encoded, so \(\gamma_{\vb{x}}\) enters as a prior probability distribution---the distribution of the object ensemble, fixed before any measurement~\cite{simoncelli2001natural,buccigrossi1999image}---with mean and covariance estimated from an image dataset~\cite{markley2024information,kabuli2026designing}. In communication the sender controls \(\gamma_{\vb{x}}\) through encoding and power allocation; when the propagation environment is fixed and cannot be designed, as in non-line-of-sight links with Rayleigh fading, in which multipath propagation renders the channel coefficients random and only their statistics are available for design~\cite{liang2004capacity,abou2001capacity,chen2004fixed,marzetta1999capacity,tse2005fundamentals}, the input distribution is the only free variable. The programs below are posed over both variables; every optimization in this work holds the input covariance fixed and acts on the permittivity alone: the design of an imaging system is a channel-design problem under the full-wave constraints of Maxwell's equations, ranging over physically realizable permittivity profiles rather than over matrices whose squared singular values obey an imposed sum budget~\cite{shlezinger2018measurement,carson2012communications}.

\section{Mutual Information of Phase Retrieval} \label{sec:phase_retrieval}
~\\ \paragraph{Channel Model and Mutual Information.}
Upon initial inspection, square-law receivers seem to limit recovery to signal amplitudes \(\abs{\vb x} \in \mathbb{R}^{M_S}_{\geq 0}\), precluding observation of the phases. 
As we demonstrate numerically below, this intuition is incomplete: when the signal phase itself carries information, an oversampled interferometric receiver makes relative phase information-bearing and, beyond the injectivity threshold, can recover the original signal up to a global phase (cf. Refs.~\cite{shechtman2015phase,shlezinger2018measurement}); the resulting mutual information exceeds what is achievable with amplitude-only recovery.

We realize our input signals \(\vb{x}\) as a distribution of free currents \(\vb{j}_i\) comprising an image in the source (also known as object) region \(S\).
The final outputs \(\vb{y}\) are realized detected field intensities, \(y_i = |(\ve_R)_i + n_i|^{2}\); 
Amplitude, or element-wise square-law detection, enacts a nonlinear operation as part of the channel
\begin{equation} \label{eq:coh_to_inc_channel_law}
    \vb{y} = |\Gt \vb{x} + \vb{n}|^{\circ2}.
\end{equation}
where \(\lvert \cdot \rvert^{\circ2}\) denotes the elementwise squared modulus.
In the scalar case, the capacity-achieving distribution for \(\vb{x}\) has previously been shown to be both discrete with an infinite number of mass points, and therefore cannot be analytically optimized~\cite{katz2004capacity}. Upon expanding Eq.~\eqref{eq:coh_to_inc_channel_law}, the output intensities at the detector experience a noise amplitude that scales with the amplitude of the field signal, leading to heteroscedastic noise, a realistic phenomenon observed in photon-number channels~\cite{wyner1988capacity,chakraborty2009poisson}. 
Expanding $|\vb{e}_R + \vb{n}|^{\circ 2} = |\vb{e}_R|^{\circ 2} +2\,\mathrm{Re}(\vb{e}_R^{*}\odot\vb{n}) + |\vb{n}|^{\circ 2}$,
the cross term has variance $\propto |\vb{e}_R|^{\circ 2}$, recovering a linear signal-dependent variance common in the literature on direct-detection channels~\cite{moser2012capacity}. 
Here this effect instead emerges from field--noise mixing under $|\cdot|^{\circ 2}$ rather than being imposed after readout.
Co-optimization of the optical front end \(\varepsilon(\vb r)\) and the sender distribution can therefore be written
\begin{equation} \label{eq:original_phase_retrieval_I}
    \begin{aligned}
        \max_{\varepsilon(\vb r), \gamma_{\vb x}} \quad & \mathcal{I}(\vb{j}_i;|\G_{t,RS}(\boldsymbol{\varepsilon})\vj_i+\vn|^{\circ2})\\ 
        \textrm{s.t.} \quad & \left[\nabla \times \nabla \times - \varepsilon(\vb r) \omega^2 \right]\G_t(\vb r, \vb r';\varepsilon) = \omega^2 \vb{I} \delta(\vb{r} - \vb{r'}) , \\ 
        & \textrm{Power constraint on } \gamma_{\vb x}.
    \end{aligned}
\end{equation}

To provide a more tractable, albeit non-capacity-achieving, value of mutual information, we focus on probability densities for \(\vj_i, \vb{e}_R\) that are zero-mean circularly-symmetric complex Gaussian.
Because the channel capacity is a supremum over input distributions, the mutual information evaluated under this Gaussian restriction is a lower bound on the capacity of the intensity-detection channel, whose capacity-achieving input is not Gaussian and, in the scalar case, is discrete~\cite{katz2004capacity}. The restriction reduces the optimization over input distributions to an optimization over the covariance $\vb{Q}$, and is the restriction adopted in mutual-information-based design of measurement matrices~\cite{shlezinger2018measurement}; comparable Gaussian simplifications are standard in the information theory literature~\cite{cover2006elements,diggavi2001worst,poggiolini2012GN}. 
Regardless, the induced distribution of the intensity output signals are \textit{not} Gaussian, and the standard \(\log\det \) objective in Eq.~\eqref{eq:logdet} does not apply.
Still, by assuming \(\vj_i \sim \mathcal{CN}(0,\vb Q)\) with covariance matrix \(\vb Q\), we restrict the family of input distributions and re-write Eq.~\eqref{eq:original_phase_retrieval_I} as
\begin{equation} 
\begin{aligned}    
\label{eq:ICAWGN_general_mutual_info_bound_objective}
    \max_{\boldsymbol{\varepsilon}(\vb{r}), \vb{Q}} \quad & \mathcal{I}(\vb{j}_i;|\G_{t,RS}(\boldsymbol{\varepsilon})\vj_i+\vn|^{\circ2}) \\
    \textrm{s.t.} \quad &  \left[\nabla \times \nabla \times - \varepsilon(\vb r) \omega^2 \right]\G_t(\vb r, \vb r';\varepsilon) = \omega^2 \vb{I} \delta(\vb{r} - \vb{r'}) \\
    & \vb{j}_i \sim \mathcal{CN}(0,\vb{Q}), \, \vb n \sim \mathcal{CN}(0, N \mathbb I), \\
    & \Tr[\vb{Q}] \leq P
\end{aligned}
\end{equation}
Under AWGN and the average-power constraint \(\Tr[\vb{Q}]\leq P\), the zero-mean circularly symmetric complex Gaussian family is capacity-achieving for the coherent channel of Eq.~\eqref{eq:logdet}, with \(\vb Q\) optimized under that constraint~\cite{telatar1999capacity}. For any fixed \(\vb Q\), evaluating the mutual information of Eq.~\eqref{eq:ICAWGN_general_mutual_info_bound_objective} compares intensity detection against the coherent Gaussian-input mutual information for the same input statistics, and the difference between the two values is the information lost to square-law detection. 

To evaluate Eq.~\eqref{eq:ICAWGN_general_mutual_info_bound_objective}, we may combine the channel model Eq.~\eqref{eq:coh_to_inc_channel_law} with Eq.~\eqref{eq:mutual_info_integral_universal} (see Appendix~\ref{asec:phase_retrieval_integral}) to find
\begin{equation}
\begin{gathered}
    \mathcal{I}(\vb{x}; \vb{y})
    = \log \det\!\left(\I_{M_R} + \frac{1}{N}\Gt \vb{Q} \Gt^{\dagger}\right) + \alpha, \\
    \alpha
    \equiv \int_{\mathbb{C}^{M_S}} \int_{\mathbb{R}^{M_R}_{\geq 0}}
    \gamma_{\vb{x}} \gamma_{\vb{y}|\vb{x}}
    \left(
        \log\!\left[
            \frac{g_2(\vb{y}|\vb{x})}{g_1(\vb{y})}
        \right]
    \right)
    \, d\vb{y}\, d\vb{x}
\end{gathered}
\label{eq:mutual_information_integral_alpha}
\end{equation}
where \(g_1(\vb{y}) \equiv \det(N \I_{M_R} +\Gt\vb{Q}\Gt^{\dagger})\gamma_{\vb{y}}\) and \(g_2(\vb{y}|\vb{x}) \equiv N^{M_R} \gamma_{\vb{y}|\vb{x}}\).
By assumption \(\gamma_{\vb x}\) is Gaussian; under the channel model, \(\gamma_{\vb y|\vb x}\) is a product of noncentral chi-squared densities, and \(\gamma_{\vb y}\) has exponential marginals that are mutually independent exactly when \(\G_{t,RS}\vb{Q}\G_{t,RS}^{\dagger}\) is diagonal (see Appendix~\ref{asec:phase_retrieval_integral}).
Equation~\eqref{eq:mutual_information_integral_alpha} is the coherent log-determinant term plus \(\alpha\), a nonpositive scalar equal to the mutual information \textit{lost} to intensity detection: since the intensities are a deterministic function of the noisy fields, the data-processing inequality bounds \(\mathcal{I}(\vb{x};\vb{y})\) by the coherent value~\cite{cover2006elements}. 
In the coherent case, where the detector records the complex field rather than its intensity, both output densities are complex Gaussian and the mutual information equals the log-determinant exactly, recovering \(\alpha = 0\).
Finally, unlike the coherent case, concavity of the integral in \(\vb{Q}\) has not been established, and the waterfilling argument does not transfer even at a fixed channel matrix (Appendix~\ref{asec:phase_retrieval_integral}). Regardless, this integral is computable with Monte Carlo methods (see Appendix~\ref{asec:numerics_phase_retrieval}).

At fixed input and noise statistics, the magnitude of $\alpha$, the mutual information lost to square-law detection, is set by the amount of relative-phase information the intensity measurements retain. Whenever the front end mixes the sources, the relative phases enter the recorded intensities; recovering them additionally requires redundancy among the measurements, set by the number of detectors relative to the number of sources, and a sufficient redundancy determines the source field up to a global phase. The ratio $M_R/M_S$ therefore divides the phase-retrieval problem into two complementary regimes.

~\\ \paragraph{Receiver-Limited Regime.}
When detectors are scarce, $M_S \geq M_R$, the field-to-intensity conversion maps multiple input degrees of freedom to each detector, and the complete set of phase relationships within the source cannot be resolved~\cite{candes2013phaselift,bandeira2014saving,conca2015algebraic,vinzant2015small}. The intensity vector $\vb{y}\in\mathbb{R}^{M_R}_{\geq 0}$ exposes at most $M_R$ real degrees of freedom per channel use, each carrying at most $\tfrac{1}{2}\log(1+\mathrm{SNR})$ at high SNR~\cite{tse2005fundamentals}, so the mutual information grows at most as $\tfrac{M_R}{2}\log(\mathrm{SNR})$: one real degree of freedom per detector, half the coherent count. This bound is the vector counterpart of the classical result that amplitude modulation carries half the pre-log of phase-coherent modulation~\cite{blachman1953comparison}, and output-entropy subadditivity sharpens the count into a converse: for every front end and every Gaussian input, the mutual information is bounded by a sum of $M_R$ scalar amplitude-detection terms, with equality precisely for the decoupled channel introduced below (Appendix~\ref{asec:optimal_product_channel}). The converse fixes the pre-log and the equality case but does not order front ends with distinct per-detector powers; whether point focusing maximizes the mutual information over every front end at $M_R<M_S$, where the measured intensities remain nonlinear in the source degrees of freedom, is an open question we leave to future work.

The square law acts detector by detector, so the channel factorizes exactly when the fields at distinct detectors fluctuate independently. Across the ensemble of transmitted signals, the propagated field $\vb{e}_R=\G_{t,RS}\vb{j}_i$ has covariance $\mathbb{E}[\vb{e}_R\vb{e}_R^{\dagger}]=\G_{t,RS}\vb{Q}\G_{t,RS}^{\dagger}$, and the channel is decoupled when this covariance is diagonal in the detector basis: the fields at distinct detectors are then uncorrelated and, being jointly Gaussian, statistically independent. Equivalently, the rows of $\G_{t,RS}$ are orthogonal under the $\vb{Q}$-weighted inner product (Appendix~\ref{asec:phase_retrieval_integral}), as realized by a mode-sorting front end that routes each of $M_R$ orthogonal source modes to its own detector; we refer to this configuration as the mode-sorted channel. Under the isotropic covariance $\vb{Q}=(P/M_S)\I_{M_S}$ of our numerical results, the routed modes are right-singular vectors of $\G_{t,RS}$, received with gains equal to the associated singular values. Point focusing, in which each row of $\G_{t,RS}$ carries a single nonzero entry so that each detector collects the field of a single source, is the special case in which the sorted modes are the source points themselves; it realizes the decoupled channel when $\vb{Q}$ is diagonal in the source-point basis.

The diagonal restriction at the receiver parallels the statement of source incoherence in Sec.~\ref{sec:incoherent_MI}: source incoherence restricts the temporal coherence matrix $\langle\vb{j}_i\vb{j}_i^{\dagger}\rangle$ to its diagonal at the sender within each channel use, while the decoupled channel restricts the ensemble covariance $\mathbb{E}[\vb{e}_R\vb{e}_R^{\dagger}]$ to its diagonal at the receiver across channel uses; the two averages are distinct. The decoupled channel is the analytically solvable case of the phase-retrieval channel: the marginal output density factorizes into a product of exponential laws, the phase integral defining it admits a closed form (Appendix~\ref{asec:phase_retrieval_integral}), and the mutual information decomposes exactly into a sum of $M_R$ scalar terms. Conditioned on the transmitted signal, each detector intensity depends only on the modulus of its own mode coefficient, so the information the channel conveys passes entirely through the $M_R$ sorted-mode amplitudes, one real degree of freedom per detector.

Specializing Eq.~\eqref{eq:mutual_information_integral_alpha} to the mode-sorted channel, where $\G_{t,RS}\vb{Q}\G_{t,RS}^{\dagger}$ is diagonal, and evaluating $\alpha$ in the high-SNR limit (Appendix~\ref{asec:phase_retrieval_integral}), we obtain
\begin{equation} \label{eq:half_logdet_asymptotic_diagonal_hiSNR}
\begin{aligned}
\mathcal{I}(\vb{j}_i;\vb{y}) &= \tfrac12\,\log\det\!\Big(\I_{M_R}+\tfrac1N \G_{t,RS} \vb{Q} \G_{t,RS}^{\dagger}\Big) \\
&\quad + M_R c_0 + o(1), \\
c_0 &= \tfrac12 + \tfrac{\gamma_{E}}{2} - \tfrac12\log(4\pi),
\end{aligned}
\end{equation}
where \(\gamma_E\) is the Euler--Mascheroni constant ($\approx 0.577$), so that $c_0 \approx -0.477$ nats; the $o(1)$ remainder vanishes as the per-detector SNR grows. The leading term is exactly half the coherent log-determinant of Eq.~\eqref{eq:logdet} at the same channel and input covariance, its pre-log $M_R/2$ attains the upper bound above, and the per-detector constant $c_0$ parallels the classical scalar analysis of amplitude detection~\cite{blachman1953comparison}.

Equation~\eqref{eq:half_logdet_asymptotic_diagonal_hiSNR} fixes the mutual information of the mode-sorted channel, and the recovered degrees of freedom are the sorted-mode amplitudes rather than, in general, the source amplitudes: each sorted mode is a fixed superposition of source currents, so its modulus depends on the relative phases among the sources it combines, and a delocalized mode-sorted channel therefore conveys partial phase content. The two sets of amplitudes coincide for point focusing at $M_S=M_R$, a permutation channel in which each detector measures $|(\G_{t,RS}\vb{j}_i)_k|^{2}$ for a single source, discarding every relative phase; Eq.~\eqref{eq:half_logdet_asymptotic_diagonal_hiSNR} evaluated at such a design is the source-amplitude-only benchmark against which the interferometric structures of the following section are compared. The ratio $M_R/M_S$ determines the receiver geometry: imaging typically operates in the receiver-limited regime, with far more discernible source points than detectors; phase retrieval requires the opposite, $M_S<M_R$, where oversampling renders the source phases information-bearing and their recovery requires an interferometric front end, as we show next.

\begin{figure*}[t]
    \centering
    \includegraphics[width=\linewidth]{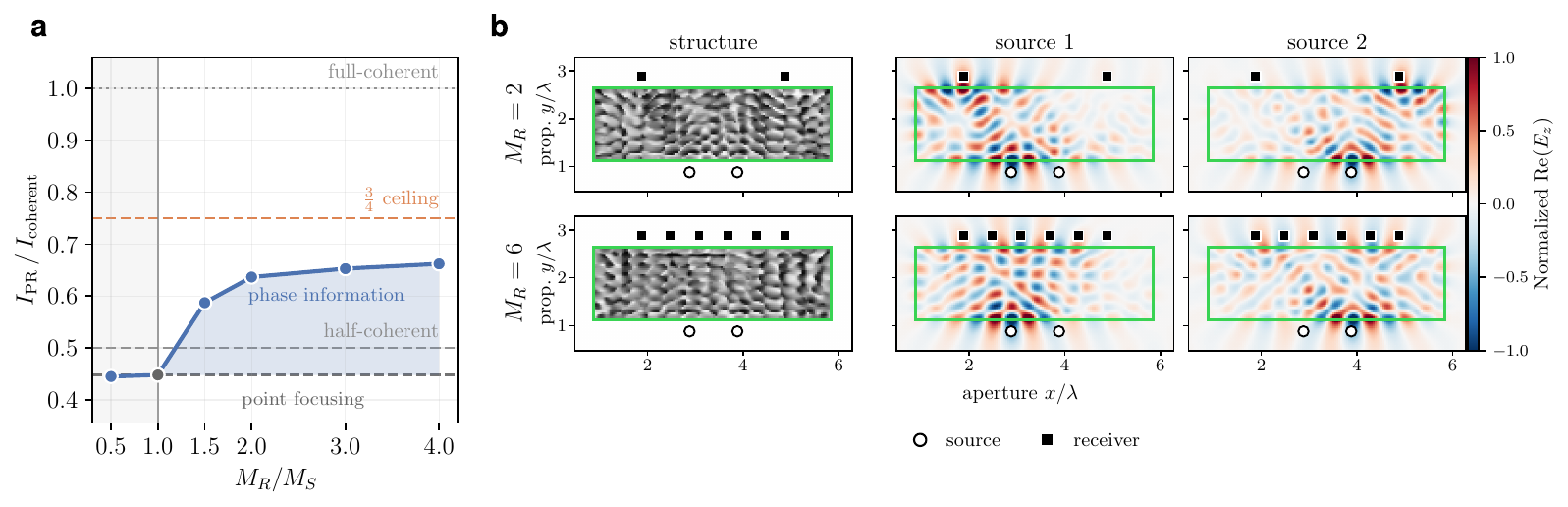}
    \caption{\textbf{Phase retrieval across the injectivity threshold at $M_S=2$ coherent sources.} (a) Mutual information of topology-optimized front ends as a function of the detector-to-source ratio $M_R/M_S$, normalized by the coherent mutual information of Eq.~\eqref{eq:logdet} on the same spatial structure and input statistics. From top to bottom, dashed lines mark the coherent value, the $(2M_S-1)/(2M_S)=3/4$ pre-log upper bound, half of the coherent value, and the amplitude-only point-focusing baseline, the optimized $M_R=M_S=2$ value, which has close numerical agreement with our derived asymptotic value in Eq.~\eqref{eq:half_logdet_asymptotic_diagonal_hiSNR}. The shaded region is the excess of the optimized value over that baseline, which we interpret as the approximate mutual information carried by the relative source phases; past the threshold this excess carries the total above the half-coherent benchmark. (b) Optimized permittivity profiles and the real part of the field radiated by each source driven alone, at $M_R=2$ (below the $4M_S-4=4$ injectivity threshold) and $M_R=6$ (above it); circles mark sources, squares mark detectors. The $M_R=2$ optimum point focuses; the $M_R=6$ optimum is interferometric. Front ends in both panels are obtained by propagating gradients through the differentiable Gaussian-matched surrogate of Appendix~\ref{asec:surrogate_objective}; the values in (a) are Monte Carlo estimates of the integral in Eq.~\eqref{eq:mutual_information_integral_alpha}.}
    \label{fig:interferometric_sources}
\end{figure*}

~\\ \paragraph{Receiver Redundancy and Phase Retrieval.}
In the opposite regime, $M_S < M_R$, the detectors outnumber the sources, and the surplus measurements make relative phase information-bearing and, beyond the injectivity threshold, recoverable up to a global phase. Each source signal is complex, so the $M_S$ sources carry $2M_S$ real unknowns per channel use. One combination of them, the global phase common to all sources, is invisible to every intensity measurement. A sufficient number of independent intensities over-determines the remaining $2M_S-1$ parameters and fixes the field up to that global phase. 
Phase-retrieval theory fixes how many measurements are sufficient. Parameter counting alone requires $M_R \geq 2M_S-1$, the real dimension of the rank-one coherence manifold, and the sharp threshold for recovering almost every field is one measurement higher: $2M_S$ suitably chosen measurements determine almost every source field, whereas $2M_S-1$ cannot~\cite{flammia2005minimal,finkelstein2004pure}. At $M_R \geq 4M_S-4$, generic measurements determine every source field~\cite{conca2015algebraic,bodmann2015stable}, and at $M_S=2$ this count is exact~\cite{bandeira2014saving, conca2015algebraic}. The mutual information quantifies what the intensities carry rather than what a tractable decoder extracts from them; in high-dimensional models with Gaussian measurement ensembles, recovery can be information-theoretically possible where no known efficient algorithm succeeds~\cite{mondelli2019fundamental,barbier2019optimal}.

As $M_R$ grows past $M_S$, the source phases become progressively information-bearing. The mutual information then rises from the amplitude-only benchmark of the receiver-limited regime towards the coherent value. The benchmark pre-log in this regime is $M_S/2$, since a diagonal $\G_{t,RS}\vb{Q}\G_{t,RS}^{\dagger}$ has rank at most $M_S$ and a decoupled front end therefore routes to at most $M_S$ detectors. The coherent value is not reached: the global phase is lost under every front end. By the parameter count, each of the $2M_S-1$ recoverable real parameters would contribute $\tfrac{1}{2}\log\mathrm{SNR}$ at high SNR, the per-degree-of-freedom growth of linear Gaussian channels~\cite{tse2005fundamentals} adopted in noncoherent degree-of-freedom analyses~\cite{durisi2011high}, placing the expected pre-log at $(2M_S-1)/2$ against the coherent $M_S$. At $M_S=2$ the count is three quarters of the coherent pre-log, the parameter-count line marked in Fig.~\ref{fig:interferometric_sources}.

Recovering the phase requires a front end that mixes the source fields, and the fields radiated by each source driven alone in Fig.~\ref{fig:interferometric_sources}(b) show this mixing directly. Expressing the source amplitudes as $r_j \ge 0$ with phases $\phi_j$, the noiseless intensity at detector $i$ is:
\begin{equation} \label{eq:cross_term_intensity}
 \sum_j \big|(\G_{t,RS})_{ij}\big|^2 r_j^2 + \sum_{j \ne k} (\G_{t,RS})_{ij}\,(\G_{t,RS})_{ik}^{*}\, r_j r_k\, e^{i(\phi_j - \phi_k)}.
\end{equation}
The relative phases $\phi_j-\phi_k$ enter only through the second sum, the cross terms between distinct sources. A front end that sends each source to its own detector annihilates the cross terms, $(\G_{t,RS})_{ij}(\G_{t,RS})_{ik}^{*}=0$ for $j\neq k$, and is therefore phase-blind; such a front end is the point-focusing design of the receiver-limited regime. Reading the phase requires the opposite: an \emph{interferometric} front end that routes several sources onto shared detectors so that their fields interfere. The interference is formed within the front end and is registered by ordinary intensity detectors, which distinguishes this arrangement from phase-sensitive detection (Sec.~\ref{sec:channel_laws}), where the detector itself measures field cross-correlations. Under the temporal average that defines source incoherence in Sec.~\ref{sec:incoherent_MI}, the cross terms of Eq.~\eqref{eq:cross_term_intensity} vanish for every front end, preventing the relative phases from entering the intensities (Appendix~\ref{asec:hadamard_product_derivation}).
 
~\\ \paragraph{Unified Coherence-Based Linear Channel Model.} The cross-term structure of Eq.~\eqref{eq:cross_term_intensity} is one instance of a general linear construction that makes the phase-retrieval channel tractable and, at the same time, connects it to the incoherent imaging channel of Sec.~\ref{sec:incoherent_MI}. The construction tracks not the source current $\vb{j}_i$, but rather its outer product---the \emph{coherence matrix}, or mutual intensity, $\vb{j}_i\vb{j}_i^{\dagger} \in \mathbb{C}^{M_S \times M_S}$. Its diagonal $\operatorname{diag}(\vb{j}_i\vb{j}_i^{\dagger}) \in \mathbb{R}^{M_S}_{\geq 0}$ holds the source intensities, and its off-diagonal entries $(\vb{j}_i\vb{j}_i^{\dagger})_{kl},\ k\neq l$, hold the mutual intensities that carry the relative phases $\phi_k - \phi_l$. The dimension-lift exchanges the $2M_S$ real degrees of freedom of $\vb{j}_i$ for a matrix invariant under the global phase $\vb{j}_i \to e^{i\theta}\vb{j}_i$. The matrix carries $2M_S-1$ real parameters, exactly the recoverable degrees of freedom; the one lost is the global phase, invisible to every intensity measurement. Propagation is linear in the current, $\vb{e}_R = \G_{t,RS}\vb{j}_i$, and linear in the coherence matrix, $\vb{e}_R\vb{e}_R^{\dagger} = \G_{t,RS}(\vb{j}_i\vb{j}_i^{\dagger})\G_{t,RS}^{\dagger}$; vectorizing,
\begin{equation} \label{eq:kronecker_lifted_field}
\operatorname{vec}\!\big(\vb{e}_R \vb{e}_R^{\dagger}\big) = \big(\G_{t,RS} \otimes \G_{t,RS}^{*}\big) \operatorname{vec}\!\big(\vb{j}_i \vb{j}_i^{\dagger}\big) \in \mathbb{C}^{M_R^2}.
\end{equation}
Square-law detection then keeps only the diagonal of $\vb{e}_R\vb{e}_R^{\dagger}$---the per-detector intensities---which a fixed selection matrix $\Sel_R \in \{0,1\}^{M_R \times M_R^2}$ extracts:
\begin{equation} \label{eq:lifted_mutual_field_sel} 
\begin{aligned}
\big|\vb{e}_R\big|^{\circ 2} &= \Sel_R \operatorname{vec}\!\big(\vb{e}_R \vb{e}_R^{\dagger}\big) \\
&= \Sel_R\big(\G_{t,RS} \otimes \G_{t,RS}^{*}\big) \operatorname{vec}\!\big(\vb{j}_i \vb{j}_i^{\dagger}\big) \in \mathbb{R}^{M_R}_{\geq 0},
\end{aligned}
\end{equation}
with $(\Sel_R)_{p,\,(a-1)M_R + b} = \delta_{p,a}\,\delta_{a,b}$, one unit entry per row, such that $\bigl(\Sel_R\operatorname{vec}(\mathbb{X})\bigr)_p = \mathbb{X}_{pp}$. The noiseless phase-retrieval channel is thus \emph{linear} in the source coherence matrix, at the cost of squaring the dimension. The information loss occurs at detection: $\Sel_R$ keeps only $M_R$ of the $M_R^2$ entries of $\operatorname{vec}(\vb{e}_R\vb{e}_R^{\dagger})$, discarding the receiver cross-coherences, so intensity measurement is compressive in the coherence domain. Mixing via $\G_{t,RS}$ makes phase retrieval possible: propagation couples the source cross-coherences $(\vb{j}_i\vb{j}_i^{\dagger})_{kl}$, producing interference on the diagonal entries of the receiver coherence matrix~\cite{candes2013phaselift,balan2006signal}.

This lifted construction follows Ref.~\cite{shlezinger2018measurement}; the identification of the lifted vectors with the physical mutual intensity is classical~\cite{goodman2015statistical,born1999principles}, and its reconstruction from intensity measurements under known propagation is the program of phase-space tomography~\cite{waller2012phase,tian2012experimental,raymer1994complex}. In our channel the noise is complex-valued, and is added \emph{before} the square-law detection rather than to the intensities afterward, the propagator is $\G_{t,RS}(\varepsilon)$, and the figure of merit is the mutual information rather than reconstruction fidelity. The selection $\Sel_R$ that measures the receiver-intensity diagonal here is the same operation that, applied to an incoherent source, produces the Hadamard channel of Sec.~\ref{sec:incoherent_MI}.

~\\ \paragraph{Photonic Inverse Design.} \label{sec:surrogate_PR}
We solve the front end optimization of Eq.~\eqref{eq:ICAWGN_general_mutual_info_bound_objective} by gradient-based photonic inverse design over physically admissible structures. The design \(\varepsilon\) shapes the channel while \(\vb{Q}\) allocates power across the source points; in the numerical results below \(\vb{Q}\) is held fixed at the isotropic \((P/M_S)\Id_{M_S}\) with power budget \(P\), and the optimization ranges over \(\varepsilon\) alone. 
Because the exact mutual information of Eq.~\eqref{eq:mutual_information_integral_alpha} has no closed form and its Monte Carlo estimator is costly and noisy to differentiate, we supply the optimization gradient through a differentiable surrogate that replaces the lifted signal and the Rician post-detection noise by Gaussians of matched covariance (Appendix~\ref{asec:surrogate_objective}). 
The surrogate supplies the optimization gradients, while every mutual-information value we report is a Monte Carlo estimate of the exact integral in Eq.~\eqref{eq:mutual_information_integral_alpha}.

~\\ \paragraph{Numerical Results.}
We sweep the detector count $M_R$ from $1$ to $8$ across the injectivity threshold at $M_S=2$ sources held fixed $1\lambda$ apart, with isotropic covariance $\vb{Q}=\Id_{M_S}$ (unit power per source, $P=M_S$) and noise level calibrated per run as $N=\Tr[\G_{t,RS}\vb{Q}\G_{t,RS}^{\dagger}]/(20\,M_R)$, holding the average per-detector pre-detection SNR at $20$; the detectors are equally spaced along a segment of length $3\lambda$ parallel to the source line, the two lines centered on a common axis, and every run shares the same design region of dimensions $1.5 \lambda \times 5\lambda$ and material susceptibility $\chi=10+0.01i$ between them. At each $M_R$ we obtain one structure by solving the surrogate program of Appendix~\ref{asec:surrogate_objective}; the reported mutual information is a Monte Carlo estimate of the exact integral in Eq.~\eqref{eq:mutual_information_integral_alpha}, normalized by the coherent mutual information of Eq.~\eqref{eq:logdet}, evaluated in closed form on the same structure and input statistics. Because the receiving segment is fixed while $M_R$ grows, the sweep varies the number of intensity samples at constant array extent.

Figure~\ref{fig:interferometric_sources}(a) shows the transition from amplitude-only readout to phase recovery as detectors are added. At \(M_R=1\), every front end realizes the decoupled channel, since \(\Gt \vb{Q} \Gt^{\dagger}\) is a scalar. The SNR calibration then fixes the normalized mutual information independent of the structure. Thus, the one-detector case becomes a structure-independent check of Eq.~\eqref{eq:half_logdet_asymptotic_diagonal_hiSNR} rather than an optimized design. The $M_R=2$ optimum point focuses (Fig.~\ref{fig:interferometric_sources}(b)), realizing the decoupled channel at its achieved per-detector gains: its Monte Carlo estimate agrees with the asymptotic value of Eq.~\eqref{eq:half_logdet_asymptotic_diagonal_hiSNR}, resolving the $M_R c_0$ offset to within the finite-SNR correction and the positive nested-estimator bias of Appendix~\ref{asec:numerics_phase_retrieval}, and sets the amplitude-only baseline of Fig.~\ref{fig:interferometric_sources}(a) at less than half of the coherent value.

Already at $M_R=3=2M_S-1$, the number of real parameters the coherence matrix carries, the optimized value exceeds both the point-focusing baseline and the half-coherent line, even though injectivity at $M_S=2$ requires four intensity measurements~\cite{bandeira2014saving} and determination of almost every source field requires $2M_S=4$ of them~\cite{flammia2005minimal,finkelstein2004pure}: the mutual information measures the statistical dependence that the cross terms of Eq.~\eqref{eq:cross_term_intensity} establish between the measured intensities and the relative phase, not the recoverability of the field, so partial phase information is present below the recovery thresholds. The optimized value then rises through the injectivity threshold $M_R\geq 4M_S-4=4$. The excess over the baseline is an estimate of the mutual information recovered from the relative source phases. The $(2M_S-1)/(2M_S)=3/4$ ceiling bounds the ratio of high-SNR slopes, not the ratio of values at finite SNR, and the moderate-SNR values of the sweep remain below it. In the coordinates of Fig.~\ref{fig:phasespace}, the sweep holds \(M_S'=2M_S-1=3\) fixed while \(M_R'=M_R\) rises from one to eight, with the onset at \(M_R=3\) crossing the parameter-matching diagonal \(M_R'=M_S'\).

Figure~\ref{fig:interferometric_sources}(b) shows the optimized structures and the field radiated by each source driven alone. At $M_R=2$, the structure routes the field of each source onto the nearest detector, with negligible field reaching the other: the cross terms of Eq.~\eqref{eq:cross_term_intensity} are suppressed, each detector reads at most the amplitude of its assigned source, and the structure realizes the point-focusing design of the receiver-limited regime. At $M_R=6$, spanning the same receiving segment, the structure spreads the field of each source across several detectors, and the field profiles of the two sources overlap on shared detectors: the cross terms of Eq.~\eqref{eq:cross_term_intensity} survive, the relative phase enters the recorded intensities, and the front end is interferometric. The change in routing is the structural counterpart of the rise of the mutual information in Fig.~\ref{fig:interferometric_sources}(a) as detectors are added.

\section{Mutual Information of Incoherent Light} \label{sec:incoherent_MI}

\begin{figure*}
    \includegraphics[width=\linewidth]{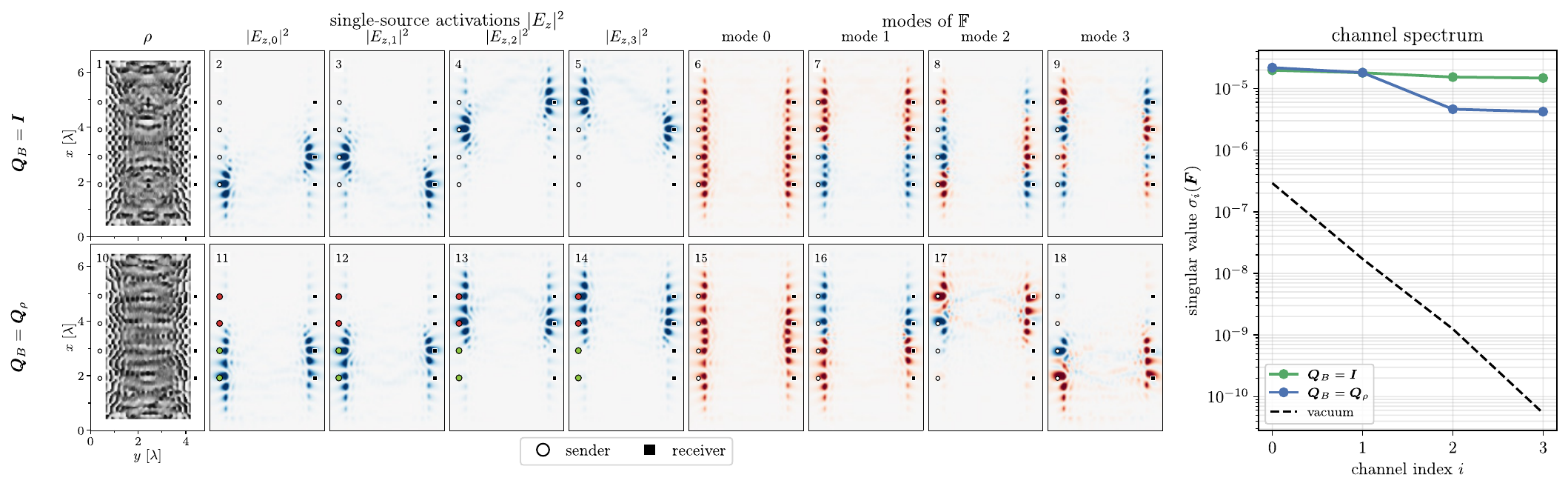}
    \caption{\textbf{Incoherent intensity channel: source correlations move the optimized front end away from point focusing.} Topology-optimized front ends maximizing the mutual information \(\mathcal{I}(\vb{B}; \vb{y}) = \frac{1}{2} \log\det \left( \I + \frac{1}{N_B} \F_{t,RS} \, \vb{Q}_B \, \F_{t,RS}^{\top} \right)\) for mutually incoherent sources read by intensity detectors at fixed geometry, under an isotropic intensity covariance \(\vb{Q}_B=\Id\) (top row) and a correlated covariance \(\vb{Q}_B=\vb{Q}_\rho\), where the sources come in strongly correlated pairs (bottom row; Appendix~\ref{asec:correlated_covariance}). The source and detector counts are held at $M_S=M_R=4$. Left: the optimized structure, the field radiated by each source driven alone, and the dominant modes of the intensity operator \(\Ft\). Circles mark sources while squares mark detectors. Red and green circles at the bottom row denote correlated pairs of sources. Right: singular values of \(\Ft\) for the two optimized structures and, for comparison, of the vacuum operator, which decay exponentially. For isotropic sources the optimized front end flattens the singular spectrum and routes each source to its own detector; for correlated sources the optimized front end departs from point focusing, distributing each source's intensity across several detectors---the two central detectors each collect intensity from two sources within a single channel use.}
    \label{fig:iso_vs_corr}
\end{figure*}

~\\ \paragraph{Channel Model.} \label{subsec:incoherent_channel_model}
The information an imaging system conveys about an object is bounded by the channel rather than by the object (Sec.~\ref{sec:intro}), and the operator that carries it depends on the coherence of the illumination~\cite{toraldo1969degrees,cox1986information,franceschetti2017wave}. The input of conventional incoherent imaging is the time-averaged object intensity \(B(\vb{r})\), the output is the time-averaged intensity recorded at the detectors, and the figure of merit is the mutual information between the two.
 Incoherent illumination means that the source currents, modeled as free currents \(\vb{j}_i(\vb{r})\), are uncorrelated at different points in the image (source, or sender region \(S\)), and their autocorrelation satisfies $\langle j_i(\vb{r}_l) j_i^{*}(\vb{r}_k) \rangle = B(\vb{r}_l) \delta(\vb{r}_l-\vb{r}_k)$.
Here, \( \langle \cdot \rangle \) denotes a temporal average, distinct from the ensemble expectation \( \mathbb{E}[\cdot ] \). The sources are quasi-monochromatic: the bandwidth is narrow against the center frequency, so a single-frequency Green's function propagates the field, while the averaging time of \(\langle\cdot\rangle\) exceeds the coherence time of the source fluctuations, so the cross terms between distinct sources vanish and their intensities add without interference~\cite{goodman2015statistical}. Spatial incoherence is a relation between sources: the field radiated by each point source holds a fixed phase relationship across the aperture, while distinct sources hold none. More generally, incoherence is a property of a basis, not exclusive to point sources: it is the statement that $\langle \vb{j}_i \vb{j}_i^{\dagger} \rangle$ is diagonal. We take the point basis here, recovering conventional incoherent imaging, but the model admits any orthogonal mode basis in which the source coherence matrix is diagonal: the same diagonal-coherence restriction that distinguishes the incoherent input from the coherent one of Sec.~\ref{sec:phase_retrieval}. In communication terms, this restriction removes the phases of \(\vj_i\) from the sender's signaling degrees of freedom: each transmitted signal (each object or image presented to the channel) undergoes the within-use temporal average \(\langle\cdot\rangle\) over its phases, such that only the intensities remain available for encoding.

As in Sec.~\ref{sec:theory}, the field incident on the receiver region \(R\) due to currents \(\vb{j}_i\) in the object region \(S\) is \( \ve_R = \G_{t,RS} \vb{j}_i\). Imposing source incoherence, \(\langle \vb{j}_i \vb{j}_i^{\dagger} \rangle = \operatorname{diag}(\vb{B})\), the time-averaged intensity at detector \(p\) is
\begin{equation}
    \avg{\big|(\ve_R)_p\big|^{2}} = \sum_{\ell=1}^{M_S} \big|(\G_{t,RS})_{p\ell}\big|^{2}\, B_\ell,
\end{equation}
the cross terms between distinct sources vanishing under \(\langle\cdot\rangle\) (Appendix~\ref{asec:hadamard_product_derivation}). Collecting the detector intensities into a vector,
\begin{equation}
    \langle \left| \vb{e}_R \right|^{\circ 2}  \rangle = \big(\G_{t,RS}^{*} \odot \G_{t,RS}\big)\vb{B}.
\end{equation}
The Hadamard square of the Green's function is the special case of the linear operator of Eq.~\eqref{eq:lifted_mutual_field_sel} obtained by right-multiplication with the source-side selection matrix \(\Sel_S^{\top}\), which enforces source incoherence (Appendix~\ref{asec:hadamard_product_derivation}).

We define \(\F_{t,RS}\) to be the linear point-spread-function (PSF) operator that maps the object intensity vector \(\vb{B}\) to the time-averaged receiver intensity vector \(\avg{|\ve_R|^{\circ 2}}\), constructed from the Green's function as:
\begin{equation}
    \F_{t,RS} := \G_{t,RS}^{*} \odot \G_{t,RS}.
\end{equation}
Each entry \((\F_{t,RS})_{p\ell} = \big|(\G_{t,RS})_{p\ell}\big|^{2}\) is the intensity response at detector \(p\) to a unit-intensity point source \(\ell\), the sampled PSF of conventional imaging; in a shift-invariant system, as in planar far-field imaging, this map reduces to a convolution of the object intensity with the PSF~\cite{goodman2015statistical}. With the PSF operator as a linear channel, the recorded output is the time-averaged intensity vector plus readout noise:
\begin{equation} \label{eq:incoh_channel_model_definitions}
    \vb{y} := \avg{ | \vb{e}_R |^{\circ 2} } + \vb{n} = \F_{t,RS} \vb{B} + \vb{n},
\end{equation}
where the input is $\vb{B} := \langle | \vb{j}_i |^{\circ 2} \rangle \sim \mathcal{N}(\boldsymbol{\mu}_{B}, \vb{Q}_B)$ and $\vb{n} \sim \mathcal{N}(\vb{0}, N_B \I)$ is real additive intensity noise. This instantiates Eq.~\eqref{eq:channel_law_universal} with input $\vb{x} \to \vb{B}$, channel $\vb{H} \to \F_{t,RS}$, and $f$ the identity, with $\mathbb{K} = \mathbb{R}$: unlike the phase-retrieval channel, where the square law makes $f$ nonlinear, source incoherence has already linearized the map by collapsing the coherence matrix to its diagonal, so the intensity channel is linear in $\vb{B}$ from the outset.

To obtain a tractable channel model, we take the source intensities $\vb B$ as multivariate Gaussian at fixed covariance $\vb Q_B$, the entropy-maximizing choice for a given second moment~\cite{cover2006elements}. 
The negative tail of the Gaussian is negligible when each mean intensity sufficiently exceeds its standard deviation, with the condition enforced by the margin constraint of Eq.~\eqref{eq:margin_constraint} below.
We also note that at low SNR the intensity noise \(\vb{n}\) can produce negative detected values: the readout $\vb{y} \sim \mathcal{N}\big(\F_{t,RS}\boldsymbol{\mu}_{B},\, N_B\I_{M_R}+\F_{t,RS} \vb{Q}_B \F_{t,RS}^\top\big) \in \mathbb{R}^{M_R}$ is a real-valued digital quantity rather than a physical nonnegative intensity, as in existing intensity-channel models~\cite{lapidoth2009capacity,li2020capacity}. 
In communication terms, the information-carrying variables are the \emph{intensity fluctuations} about a positive mean intensity, \(\vb{B} = \boldsymbol{\mu}_B + \delta\vb{B}\), such that \(\vb{Q}_B:= \operatorname{Cov}[\vb{B}]=\mathbb{E}_{\vb{B}}\big[(\delta\vb{B})(\delta \vb{B})^{\top}\big]\). That is, only the fluctuations \(\delta\vb B\) are information-carrying: the mutual information is invariant to the deterministic output shift \(\F_{t,RS}\boldsymbol{\mu}_B\), and \(\boldsymbol{\mu}_B\) affects the model only through the nonnegativity margin below.
In imaging applications, \(\boldsymbol{\mu}_B\) represents the mean intensity, or average brightness, of a given source object distribution or image dataset, and the diagonal elements of \(\vb{Q}_B\) represent contrast values, while the off-diagonal elements represent correlations across the ensemble. 

Equation~\eqref{eq:incoh_channel_model_definitions} defines a linear AWGN channel with real Gaussian input and noise, and therefore its mutual information is~\cite{cover2006elements}:
\begin{equation} \label{eq:incoh_channel_MI}
    \mathcal{I}(\vb{B}; \vb{y}) = \frac{1}{2} \log\det \left( \I + \frac{1}{N_B} \F_{t,RS} \, \vb{Q}_B \, \F_{t,RS}^{\top} \right).
\end{equation}
For a fixed source covariance $\vb{Q}_B$, Eq.~\eqref{eq:incoh_channel_MI} is maximized over the channel by the standard MIMO alignment argument~\cite{carson2012communications,telatar1999capacity}: writing $\Ft = U_{\F}\Sigma_{\F}V_{\F}^{\top}$ and $\vb{Q}_B = V_Q\Lambda_Q V_Q^{\top}$, the mutual information is largest when the channel's right-singular basis $V_{\F}$ is aligned with the covariance eigenbasis $V_Q$, so that the strongest channel modes carry the highest-variance source directions. For the intensity channel, this alignment is not always attainable. $\Ft$ is not a free operator: it must equal the Hadamard square of a physical Green's function $\G_{t,RS}$. That constraint couples the singular vectors $V_{\F}$ and the singular values $\Sigma_{\F}$, so the two cannot be chosen independently to meet the alignment. More fundamentally, the intensity mutual information depends on $\G_{t,RS}$ only through its Hadamard square $\Ft = \G_{t,RS}^{*}\odot\G_{t,RS}$. The favorable Green's functions are then those whose $\Ft$ has a favorable singular spectrum, not those whose own singular spectrum is optimal for the coherent channel. The intensity-optimal and coherent-optimal front ends are therefore generally distinct.

The co-optimization of the Gaussian input distribution and the structure may be written as:
\begin{subequations} \label{eq:Shannon_GaussianTO}
\begin{align}
         \max_{\varepsilon(\vb{r}), \boldsymbol{\mu}_B, \vb{Q}_B} \;\,  & \frac{1}{2}\log \det( \I + \frac{1}{N_B} \F_{t,RS}(\varepsilon) \, \vb{Q}_B \, \F_{t,RS}^{\top}(\varepsilon)) \label{eq:shannon_TO_objective} \\
        \text{s.t.} \;\, & \, \F_{t,RS}(\varepsilon) = \G_{t,RS}^*(\varepsilon) \odot \G_{t,RS}(\varepsilon) \label{eq:shannon_TO_objective_power} \\
        & \vb{Q}_B \succeq \vb{0} \\
        & (\curl\curl - \varepsilon(\vb{r})\omega^2) \G_t(\vb{r},\vb{r}';\varepsilon) = \omega^2 \vb{I} \delta(\vb{r}-\vb{r}') \\
        & \sqrt{(\vb{Q}_B)_{nn}} \leq \kappa^{-1} (\boldsymbol{\mu}_B)_n \quad \forall \; n \in \{1,2,\ldots,M_S\} \label{eq:margin_constraint} \\
        &\vb{1}^{\top} \boldsymbol{\mu}_B \leq P 
\end{align}
\end{subequations}
where \(P\) is a power budget on the total mean source intensity and \(\kappa\) is a nonnegativity margin: constraint~\eqref{eq:margin_constraint} holds each mean intensity at least \(\kappa\) standard deviations above zero, and setting \(\kappa=\sqrt{2\ln(M_S/\delta)}\) holds the probability that any of the \(M_S\) components is negative below a prescribed tolerance \(\delta\), by the Gaussian tail bound and a union bound~\cite{vershynin2018high}. As before, optimizing an imaging front end entails optimizing over \(\varepsilon(\vb r)\) at fixed \(\vb Q_B\); we do not optimize over \(\vb Q_B\) in this work.

This constraint restricts the input beyond the standard MIMO trace bound on $\vb{Q}_B$, and under it the Gaussian input is not capacity-achieving~\cite{lapidoth2009capacity,dytso2019amplitude}. In imaging, $\vb{Q}_B$ is fixed by the source ensemble and constraint~\eqref{eq:margin_constraint} is a condition on the model rather than on the design, holding whenever each mean intensity exceeds its standard deviation by the factor $\kappa$; in communication, where $\boldsymbol{\mu}_B$ and $\vb{Q}_B$ are themselves design variables, the constraint binds at the optimum, since the mutual information increases with the fluctuation power at every fixed mean. The margin is independent of the noise level: it fixes the standard-deviation-to-mean ratio of the input, not its relation to $N_B$, so the mutual information of Eq.~\eqref{eq:incoh_channel_MI} is exact at every SNR within the model of Eq.~\eqref{eq:incoh_channel_model_definitions}.

~\\ \paragraph{Conditions for MI-optimality of Point Focusing.} In the case of equal source and detector counts (\(M_R=M_S=M\)) and an isotropic covariance matrix \(\vb{Q}_B=q_B\I_{M}\) with \(q_B>0\), point focusing is MI-optimal whenever structuring can flatten the singular value spectrum of \(\F_{t,RS}\) without reduction of the sum of its squared singular values. For general-purpose imaging, where the orientation and correlations of the source are unknown to the designer, the isotropic covariance is the Gaussian model that treats every source direction identically.

Under these assumptions, the mutual information of Eq.~\eqref{eq:incoh_channel_MI} depends on \(\F_{t,RS}\) only through its singular values. Writing the singular value decomposition \(\F_{t,RS}=U_{\F}\Sigma_{\F}V_{\F}^{\top}\)
with ordered singular values \(\sigma_{\F,1}\geq\cdots\geq\sigma_{\F,M}\geq 0\), and using \(V_\F^{\top}\vb{Q}_B V_\F=q_B\I_M\), the resulting covariance is independent of the right singular basis: \(\F_{t,RS}\vb{Q}_B\F_{t,RS}^{\top}=q_B\,U_{\F}\Sigma_{\F}^{2}U_{\F}^{\top}\). 
Furthermore, we note that the determinant is invariant under the orthogonal similarity by \(U_{\F}\), so Eq.~\eqref{eq:incoh_channel_MI} reduces to \(\mathcal{I}(\vb{B};\vb{y})= \frac{1}{2}\sum_{i=1}^{M}\log\big(1+\tfrac{q_B}{N_B}\,\sigma_{\F,i}^{2}\big)\).

Because \(t\mapsto\log\big(1+\tfrac{q_B}{N_B}t\big)\) is strictly concave, Jensen's inequality~\cite{cover2006elements}, applied with equal weights \(1/M\) to the squared singular values, yields, for every channel, 
\begin{equation}
\mathcal{I}=\tfrac{1}{2}\sum_{i=1}^{M}\log\big(1+\tfrac{q_B}{N_B}\,\sigma_{\F,i}^{2}\big)\leq\tfrac{M}{2}\log\big(1+\tfrac{q_B}{N_B}\tfrac{1}{M}\norm{\F_{t,RS}}_{F}^{2}\big),
\end{equation}
where \(\norm{\F_{t,RS}}_{F}^{2}=\sum_{i,j}(\F_{t,RS})_{ij}^{2}=\sum_{i}\sigma_{\F,i}^{2}\) is the squared Frobenius norm. Equality holds when \(\sigma_{\F,i}^{2}=\norm{\F_{t,RS}}_{F}^{2}/M\) for every \(i\): the flat spectrum attains the bound by substitution, and strict concavity forces every equality case to be flat~\cite{cover2006elements}. Among channels sharing the total \(\sum_{i}\sigma_{\F,i}^{2}\), the flat spectrum therefore uniquely maximizes Eq.~\eqref{eq:incoh_channel_MI} at every noise variance \(N_B\) and thus at every SNR; when structuring can flatten the spectrum of \(\F_{t,RS}\) while preserving this total, the flat design attains the upper bound of every competitor and is optimal over the achievable set. 

Since \((\F_{t,RS})_{p\ell}=|(\G_{t,RS})_{p\ell}|^{2}\geq0\) and a flat positive spectrum yields \(\F_{t,RS}^{\top}\F_{t,RS}=\sigma^{2}\I_M\), one has \(\F_{t,RS}^{-1}=\sigma^{-2}\F_{t,RS}^{\top}\geq0\). Hence \(\F_{t,RS}\) is monomial, i.e., it has exactly one positive entry in each row and column~\cite{berman1994nonnegative}. The Gram identity then forces every nonzero entry to equal \(\sigma\), and therefore \(\F_{t,RS}=\sigma\Pi\), where \(\Pi\) is a permutation matrix. This equal-gain source--detector pairing is the structure termed generalized focusing in Ref.~\cite{kienesberger2026end}. Appendix~\ref{asec:farfield_focusing} extends this result to the radiative far field, where a unitary focusing front end preserves the supported field-channel singular values while localizing the modes into diffraction-limited source and detector cells; on this lattice, the resulting intensity operator is a scaled permutation that maximizes the mutual information under isotropic source statistics, thereby recovering conventional far-field imaging.

This agreement with conventional imaging is conditional on the source statistics rather than on the imaging geometry: it holds when the source intensities are uncorrelated and of equal variance, and the optimality of focusing follows from that assumption. Because the isotropic model treats every source direction identically, a front end designed under it is robust to the orientation of an unknown scene; the same property means the design forgoes the mutual information available to a front end tailored to a known source model, and practical source ensembles generally have correlated, unequal-variance intensity fluctuations~\cite{simoncelli2001natural}. An isotropic covariance is a reasonable default for a non-adaptive imager, but it does not exploit correlations that a known or measured source ensemble would contain.

~\\ \paragraph{Spectral Bounds on the Point-Spread-Function Operator.} The mutual information of Eq.~\eqref{eq:incoh_channel_MI} is bounded for every source covariance through the singular values of \(\F_{t,RS}\): the covariance sets the power carried on each singular mode of the channel and cannot increase their number. Given the ordered singular values \(\sigma_{\F,1}\geq\cdots\geq\sigma_{\F,M}\geq 0\) of \(\F_{t,RS}\), the left singular basis factors out of the determinant by orthogonal similarity, the rotated covariance \(V_{\F}^{\top}\vb{Q}_B V_{\F}\) inherits positive semidefiniteness from \(\vb{Q}_B\), and Hadamard's inequality---the determinant of a positive semidefinite matrix is at most the product of its diagonal entries~\cite{cover2006elements}---yields
\begin{equation}\label{eq:hadamard_diag_bound}
\begin{aligned}
&\frac{1}{2}\log\det\!\left(\I_{M}+\frac{1}{N_B}\,\F_{t,RS}\,\vb{Q}_B\,\F_{t,RS}^{\top}\right)\\
&\qquad\leq\;\frac{1}{2}\sum_{j=1}^{M}\log\!\left(1+\frac{\sigma_{\F,j}^{2}\,\big(V_{\F}^{\top}\vb{Q}_B V_{\F}\big)_{jj}}{N_B}\right),
\end{aligned}
\end{equation}
for every positive semidefinite \(\vb{Q}_B\), with equality when \(V_{\F}^{\top}\vb{Q}_B V_{\F}\) is diagonal, the alignment condition discussed above; an isotropic covariance satisfies the alignment identically. Bounds on the intensity spectrum therefore constrain the mutual information at any fixed covariance, and the question becomes how the spectrum of \(\F_{t,RS}\) is constrained by that of the field Green's function \(\G_{t,RS}\), written \(\sigma_{\G,1}\geq\sigma_{\G,2}\geq\cdots\), for which structure-agnostic limits over all permittivity profiles can be evaluated~\cite{amaolo2026maximum,virally2026many}.

We derive these bounds from the coherence-matrix construction of Sec.~\ref{sec:phase_retrieval}. Source incoherence restricts the source coherence matrix to its diagonal, an embedding by \(\Sel_S^{\top}\), and square-law detection retains the diagonal of the receiver coherence matrix, a selection by \(\Sel_R\) (Eq.~\eqref{eq:lifted_mutual_field_sel}, Appendix~\ref{asec:hadamard_product_derivation}), so the point-spread-function operator is the coherence-propagation operator of Eq.~\eqref{eq:kronecker_lifted_field} compressed on both sides:
\begin{equation}\label{eq:F_two_sided_selection}
\F_{t,RS}=\Sel_R\big(\G_{t,RS} \otimes \G_{t,RS}^{*}\big) \Sel_S^{\top}.
\end{equation}
The rows of \(\Sel_R\) and of \(\Sel_S\) are orthonormal, so the map \(\mathbb{X}\mapsto\Sel_R\mathbb{X}\Sel_S^{\top}\) restricts \(\mathbb{X}\) to a pair of subspaces and does not increase any ordered singular value, while the singular values of the Kronecker product are the pairwise products \(\sigma_{\G,i}\,\sigma_{\G,l}\)~\cite{horn1991topics}. Each intensity singular value is therefore bounded by the corresponding ordered pairwise product of coherent singular values; in particular \(\sigma_{\F,1}\leq\sigma_{\G,1}^{2}\) and \(\operatorname{rank} \big[\F_{t,RS}\big]\leq \big(\operatorname{rank}\big[\G_{t,RS}\big]\big)^{2}\).

An alternative set of bounds follows from a basic inequality for the singular values of a Hadamard product~\cite{ando1987singular,horn1991topics}: for every truncation order \(m\),
\begin{equation}\label{eq:F_kyfan_majorization}
\sum_{j=1}^{m}\sigma_{\F,j}\;\leq\;\sum_{j=1}^{m}\sigma_{\G,j}^{2},\qquad 1\leq m\leq M:
\end{equation}
the intensity spectrum is weakly majorized by the squared coherent spectrum: every partial sum of the former, in decreasing order, is bounded by the corresponding partial sum of the latter. The right-hand side is, up to the fixed factor \(1/\omega^{2}\), the coherent spectral sum bounded over all permittivity profiles in Refs.~\cite{amaolo2026maximum,virally2026many}, which limit the bare Green's-function spectrum \(\sigma_{\tilde\G,j}=\omega\,\sigma_{\G,j}\); every such limit transfers to the intensity spectrum and, through Eq.~\eqref{eq:hadamard_diag_bound}, to the mutual information at every source covariance. Equality holds at every order whenever \(\G_{t,RS}\) carries at most one nonzero entry in each row and each column---the point-focusing channels, of which the scaled permutations above are the equal-gain case---since then \(\sigma_{\F,j}=\sigma_{\G,j}^{2}\) for every \(j\). Two structures with the same coherent spectrum can realize opposite extremes of the intensity spectrum: the identity and the unitary discrete Fourier transform matrix share the flat spectrum \(\sigma_{\G,j}=1\), yet their Hadamard squares are \(\I_{M}\), with a flat intensity spectrum, and the rank-one \(M^{-1}\mathbf{1}\mathbf{1}^{\top}\). The entrywise phases separating the two are invisible to any function of \(\sigma_{\G}\). Equation~\eqref{eq:F_kyfan_majorization} is a bound over the design space at fixed coherent spectrum, attained by focusing, rather than an estimate for a given structure.

The spectral constraints combine with Eq.~\eqref{eq:hadamard_diag_bound} into a closed-form upper bound at arbitrary source covariance. Each diagonal entry \(\big(V_{\F}^{\top}\vb{Q}_B V_{\F}\big)_{jj}\) is a quadratic form of \(\vb{Q}_B\) at a unit vector and is at most the largest covariance eigenvalue \(\lambda_{\max}(\vb{Q}_B)\). The summand in Eq.~\eqref{eq:hadamard_diag_bound} is concave in \(\sigma_{\F,j}^{2}\), so Jensen's inequality with equal weights bounds the sum by its value at the flat spectrum~\cite{cover2006elements}. The total obeys \(\sum_{j}\sigma_{\F,j}^{2}\leq\sigma_{\F,1}\sum_{j}\sigma_{\F,j}\leq\sigma_{\G,1}^{2}\sum_{j}\sigma_{\G,j}^{2}\), by the compression bound above and Eq.~\eqref{eq:F_kyfan_majorization} at \(m=M\). Together,
\begin{equation}\label{eq:F_MI_ceiling}
\mathcal{I}\big(\vb{B};\vb{y}\big)\;\leq\;
\frac{M}{2}\log\!\left(1+\frac{\lambda_{\max}(\vb{Q}_B)}{N_B}\,\frac{\sigma_{\G,1}^{2}}{M}\sum_{j=1}^{M}\sigma_{\G,j}^{2}\right),
\end{equation}
for every positive semidefinite \(\vb{Q}_B\) and every noise variance \(N_B\), with equality exactly when \(\vb{Q}_B\) is isotropic and \(\F_{t,RS}\) is a scaled permutation: the point-focusing optimum established above is the unique configuration that saturates the general bound. Because the right-hand side depends on the structure only through the coherent singular values, any structure-agnostic limit on \(\sigma_{\G}\)~\cite{amaolo2026maximum,virally2026many} converts Eq.~\eqref{eq:F_MI_ceiling} into a bound on the incoherent mutual information over every photonic front end; the evaluation of these bounds for specific source--detector geometries is left to future work. 

~\\ \paragraph{Numerical Results.}
We optimize the front end for $M_S=4$ mutually incoherent sources at fixed pitch $1\,\lambda$ (a TM-polarized source line of total length $3\,\lambda$), read by $M_R=4$ intensity detectors equally spaced along a parallel segment of length $3\,\lambda$ set a distance $4\,\lambda$ away, under the two intensity covariances of Fig.~\ref{fig:iso_vs_corr}: the isotropic $\vb{Q}_B=\Id$ and the correlated $\vb{Q}_B=\vb{Q}_\rho$ defined in Appendix~\ref{asec:correlated_covariance}, which correlates the sources in adjacent pairs with coefficient $\rho=0.95$ (condition number $39$). 
Both covariances are normalized to $\Tr[\vb{Q}_B] = M_S$, so the two runs have the same total variance of the source-intensity fluctuations and differ only in how those fluctuations are correlated. Both share a design region of $6\,\lambda \times 3.5\,\lambda$ (transverse $\times$ propagation), centered in the source--detector gap with a $0.25\,\lambda$ vacuum standoff on either side, and a material susceptibility $\chi = 10 + 0.05\,i$.

Figure~\ref{fig:iso_vs_corr} shows, for each covariance, the optimized structure, the field radiated by each source driven alone, the dominant modes of \(\F_{t,RS}\), and its singular values against those of the vacuum operator, which decay exponentially. Under the isotropic covariance the optimized structure approximately flattens the singular spectrum and routes each source to its own detector, realizing the scaled-permutation optimum derived above. Under the correlated covariance the optimized structure departs from point focusing: each source's intensity is distributed across several detectors, and the two central detectors each collect intensity from two sources within a single channel use. Geometry, detector count, and material budget are identical between the two cases, so the change in routing follows from the source statistics alone. For this channel the coordinates of Fig.~\ref{fig:phasespace} follow from the counts with one correction. \(\vb{Q}_B\) is full rank and the optimized \(\F_{t,RS}\) is well conditioned in both cases, so \(M_R'=M_R=4\), and \(M_S'=M_S=4\) for the isotropic ensemble; the correlations of \(\vb{Q}_\rho\) reduce the effective source dimension below four because its covariance eigenvalues decay~(Sec.~\ref{sec:intro}). The correlated ensemble is therefore to the left of the \(M_S'=M_R'\) point at unchanged hardware.

\section{Concluding Remarks}
This work casts imaging, sensing, and communication with light of varying coherence as instances of one channel construction, in which the phase coherence of the source and the phase sensitivity of the detector select the channel law. Both intensity-detection cases reduce in the absence of noise to exact linear models in which the source coherence matrix, rather than the field, is information-carrying: selecting the receiver-intensity diagonal of the propagated coherence matrix produces the compressive phase-retrieval channel of Sec.~\ref{sec:phase_retrieval}, and restricting the source coherence matrix to its diagonal produces the Hadamard intensity channel of Sec.~\ref{sec:incoherent_MI}, so phase retrieval and incoherent imaging are two selections applied to one lifted linear map. The design variable is the permittivity profile, entering through a Maxwell-constrained Green's function with the input statistics held fixed; the mutual information is therefore optimized over physically realizable front ends rather than over matrices with normalized singular-value budgets~\cite{shlezinger2018measurement,carson2012communications}, extending the coherent-detection capacity framework of Ref.~\cite{amaolo2026maximum} to square-law detection and incoherent sources. Where classical treatments of optical and wireless information inhabit the vacuum far field or statistical non-line-of-sight abstractions---a fixed point-spread function, a plane-wave decomposition, or a channel matrix drawn from an ensemble rather than derived---the present framework derives the channel from Maxwell's equations in full generality, near field and far field, vacuum and structured media, so that evanescent components, photonic resonances, and integrated structures enter the information account on the same footing as radiative modes, with structure-agnostic bounds on what any design can achieve.

For coherent sources measured by square-law detectors, the ratio of detectors to sources separates two regimes. When detectors do not outnumber sources, each detector supplies one real degree of freedom per channel use; for $M_R = M_S$ and a mode-sorted channel, Appendix~\ref{asec:phase_retrieval_integral} establishes the high-SNR mutual information as half the coherent log-determinant plus a constant per detector---a pre-log that attains the ceiling bounding every front end under Gaussian inputs (Appendix~\ref{asec:optimal_product_channel})---in structural agreement with prior scalar analyses of amplitude detection~\cite{blachman1953comparison}. When detectors outnumber sources, as in oversampled phase-retrieval practice~\cite{shechtman2015phase}, the intensity measurements determine the field up to a global phase past the injectivity thresholds (invoked in Sec.~\ref{sec:phase_retrieval})~\cite{bodmann2015stable,conca2015algebraic,balan2006signal}, and the relative phases enter the recorded intensities only through cross terms between sources that share a detector, so their recovery requires an interferometric front end. Inverse design realizes this gain: the optimized structures are interferometric, and their Monte Carlo-estimated mutual information exceeds the point-focusing value that recovers the source amplitudes alone (Fig.~\ref{fig:interferometric_sources}).

For incoherent sources, the optimality of point focusing is conditional on the source statistics rather than the imaging geometry: focusing maximizes the mutual information at every SNR under an isotropic source covariance whenever structuring can flatten the intensity spectrum at fixed Frobenius norm (Sec.~\ref{sec:incoherent_MI}), the isotropy itself encoding the rotational symmetry assumed for an unknown scene; in the radiative far field this optimum reproduces the numerical-aperture resolution limit of conventional imaging (Appendix~\ref{asec:farfield_focusing}). Inverse design realizes both regimes (Fig.~\ref{fig:iso_vs_corr}): under an isotropic covariance the optimized structure flattens the intensity spectrum and routes each source to its own detector, while under a correlated covariance (\(\vb{Q}_B\) non-diagonal in the basis of mutual incoherence), of the kind standard in computational imaging~\cite{kabuli2026designing,markley2024information}, the optimized front end departs from point focusing, distributing each source across several detectors. The singular-value bounds developed for the Green's function in Refs.~\cite{amaolo2026maximum,virally2026many}, which constrain the achievable channel spectrum over every photonic structure, transfer to the Hadamard operator of Sec.~\ref{sec:incoherent_MI} through the majorization of Eq.~\eqref{eq:F_kyfan_majorization}, bounding the incoherent mutual information independently of the structure (Eq.~\eqref{eq:F_MI_ceiling}); the evaluation of these bounds for specific source--detector geometries is left to future work.

These results connect lines of work treated under separate restrictions. Capacity analyses of optical intensity channels treat a scalar link or a fixed transfer matrix, with real-valued noise added to the detected intensities~\cite{lapidoth2009capacity,hranilovic2006pixelated,li2020capacity,moser2012capacity,mecozzi2018information}; here the channel is the design variable, and the noise is complex field noise entering before the square law, so the signal-dependent intensity noise of direct-detection models~\cite{wyner1988capacity,chakraborty2009poisson,moser2012capacity} is emergent rather than imposed at readout. Photonic inverse design with spatially incoherent sources has been made tractable through trace formulations of ensemble-averaged power objectives~\cite{yao2022trace}; the present work instead takes the mutual information of the lifted channel as the design objective. Estimation-theoretic analyses of two-point resolution show an analogous dependence on measurement domain and source coherence: Fourier-plane intensity measurements attain the quantum Fisher-information limit for fully coherent sources and remain informative for partially coherent sources~\cite{wadood2025super}.
Classical accounts of the information content of an optical system count transmitted degrees of freedom under equal per-mode signal-to-noise ratio~\cite{toraldo1969degrees,cox1986information,franceschetti2017wave}, and rigorous electromagnetic treatments establish the decaying singular spectra that bound the mode count~\cite{piestun2000electromagnetic,miller2019waves,miller2000communicating,kuang2025bounds,gustafsson2026spatial}; the present framework retains the unequal spectra and formulates joint optimization of the channel spectrum and power allocation, while the numerical studies here optimize the structure at fixed input covariance, following the full-wave coherent-detection analyses of Refs.~\cite{amaolo2026maximum,molesky2022t}. Information-theoretic objectives have likewise guided end-to-end design of imaging front ends~\cite{lin2021end,lin2022end,markley2024information,kabuli2026designing}.

Throughout this work we restrict attention to vector channels with Gaussian input distributions, whose higher-order moments are fixed by the mean and the covariance. The intensity statistics of natural image ensembles depart from this Gaussian form: the single-pixel intensity marginal densities are asymmetric and heavy-tailed~\cite{huang1999statistics}, the intensities are strongly correlated across pixels~\cite{simoncelli2001natural}, and each realization is sparse in a transform basis~\cite{field1987relations,candes2006near}; the correlations enter the present framework through the covariance, but the heavy tails and the sparsity are properties of the distribution beyond its first two moments, so a faithful prior distribution for natural-image ensembles must be specified beyond the covariance alone. Under a non-Gaussian source the mutual information is no longer the Gaussian log-determinant of Eq.~\eqref{eq:incoh_channel_MI} and is not fixed by the first two moments $\boldsymbol{\mu}_B$ and $\vb{Q}_B$ alone, so the channel optimization does not reduce to a tractable function of the front end and the input covariance. The Gaussian case instead provides an upper bound on the mutual information of a non-Gaussian source of the same covariance~\cite{telatar1999capacity}.

Phase access at the source and at the detector admits four combinations. The coherent-to-coherent case is treated in Ref.~\cite{amaolo2026maximum}, with the two intensity-detection cases here; the fourth, incoherent sources measured by a phase-sensitive detector, is left to future work: its gain over intensity detection resides in the off-diagonal mutual intensities at the receiver, whose recovery requires measuring field cross-correlations rather than per-point intensities~\cite{goodman2015statistical,brady2025interferometric}. The treated cases also occupy the fully coherent and fully incoherent limits of source coherence; partial phase access at the source or the detector, for example a source of partial local coherence whose $\langle \vb{j}_i \vb{j}_i^{\dagger} \rangle$ is not diagonal but Toeplitz, is likewise left to future work. 

Read against Fig.~\ref{fig:phasespace}, these results classify the phase space region by region. On the diagonal $M_S'=M_R'$ at matched counts, isotropic incoherent statistics select point focusing at every SNR under the flattening condition at a fixed Frobenius norm; along the phase-retrieval column at \(M_S=2\), the optimized front ends pass from focusing at \(M_R=M_S\) to interferometric mixing from \(M_R=3\) onward, the recovered fraction of the coherent mutual information rising from below one half toward the \((2M_S-1)/(2M_S)\) parameter-count line; at unchanged hardware, correlated incoherent ensembles move the optimized front ends away from focusing (Fig.~\ref{fig:iso_vs_corr}); and the singular-value bounds of Sec.~\ref{sec:incoherent_MI} limit the mutual information of the spatially incoherent Hadamard channel at every source covariance. The classification is consistent with the counting boundary: every channel treated here shows pre-log at most \(\tfrac{1}{2}\min(M_S',M_R')\) under the Gaussian models, proven for the linear channels and consistent with the numerical slopes for phase retrieval---yet the boundary nowhere fixes the optimum. Several regions remain open. When detectors do not outnumber coherent sources, the optimal front end is unknown; Eq.~\eqref{eq:half_logdet_asymptotic_diagonal_hiSNR} gives only the high-SNR mutual information of the mode-sorted channel. The partial-coherence band and the phase-sensitive-detection row of the plane remain untreated. In the near field, evanescent access enlarges the measurement dimension, but the achievable intensity spectrum is graded rather than flat, and the optimality of focusing holds only when structuring can flatten it. An imager's position on the plane does not determine the optimal structure; the projection (where phase is discarded, together with the statistics of the source) does.

~\\ \paragraph{Outlook.}
Every channel considered in this work is static, with fixed permittivity profiles, source covariances, and measurement schemes. Depending on the design, an instrument may tune its illumination, integration time, interferometric phase, position and orientation, or, through reconfigurable or phase-change media, the structure itself. The averaging window measured against the  coherence time of the sources selects the channel law, motivating an extension of the framework to the intermediate regimes between full coherence and incoherence. When these parameters are varied between measurements, the object of the theory becomes the mutual information of a sequence, $\mathcal{I}(\vb{x};\vb{y}_1,\ldots,\vb{y}_T)$, each setting chosen by a policy informed by the preceding outcomes. For a linear Gaussian channel the posterior covariance does not depend on the observed data and the optimal sequence can be scheduled in advance; here, square-law readout, non-Gaussian scene statistics, and priors estimated during acquisition make each record informative about the subsequent setting. Reference~\cite{keshvari2026adaptive} co-designs sensor geometry, measurement policy, and inference to exceed non-adaptive information limits; the channel laws derived here supply the per-step physics for that program. The structures best suited to adaptive measurement are not necessarily the static optima established above: the optimal front end is set by the source ensemble~\cite{kabuli2026designing,markley2024information}; a prior that shifts with the object class therefore requires front ends that re-estimate ensemble parameters and reconfigure dynamically. When the scene evolves and $\vb{B}$ becomes a stochastic process, the integration time that enforces incoherence trades noise suppression against temporal bandwidth, and the figure of merit becomes an information rate. Control and adaptation extend the question of this work from a single measurement to a sequence: \textit{which reconfigurable geometry, policy, and prior jointly maximize the information rate of an instrument?}

~\\ \paragraph{Acknowledgments:} We thank Pramod Viswanath and David Brady for useful discussions. 
A portion of the simulations presented in this article were performed on computational resources managed and supported by Princeton University’s Research Computing.

~\\ \paragraph{Funding:} F.J.C., A.A., and A.W.R. acknowledge the support by a Princeton SEAS Innovation Grant. Z.L. acknowledges support from U.S. Army Research Office (award numbers W911NF2410390 and W911NF2510113). S.M. acknowledges financial support from NSERC under Discovery Grant RGPIN-2023-05818, as well as additional benefits provided by affiliations to the Regroupement Québécois sur les Matériaux de Pointe, \url{https://doi.org/10.69777/309032}, the IVADO Research Consortium, and the Lassonde Deeptech Institute. 

~\\ \paragraph{Data Availability:}
The code and datasets for fully reproducing the numerical results in this article are available on GitHub: \url{https://github.com/physical-design-bounds/incoherent-mutual-information}.

\appendix

\pagebreak

\section{The lifted coherence-matrix channel and modal alignment}\label{asec:app_preliminaries}
\subsection{Derivation of Hadamard product channel from intensity-detection compression of mutual intensity}\label{asec:hadamard_product_derivation}

Square-law detection records the diagonal entries of the receiver mutual intensity matrix, while linear propagation is bilinear in the field and carries the entire source mutual intensity to the receiver, $\avg{\ve_R\ve_R^{\dagger}}=\Gt\,\avg{\vj_i\vj_i^{\dagger}}\,\Gt^{\dagger}$: even a diagonal (incoherent) source mutual intensity produces off-diagonal mutual-coherence entries in $\avg{\ve_R\ve_R^{\dagger}}$ through the coupling in $\Gt$ by the van Cittert--Zernike theorem (usually applied in the far field)~\cite{goodman2015statistical,born1999principles}. The main text gives the dimension-lifted propagation of Eq.~\eqref{eq:kronecker_lifted_field} and the receiver selection $\Sel_R$ of Eq.~\eqref{eq:lifted_mutual_field_sel}, and states the incoherent point-spread-function operator as the two-sided compression of Eq.~\eqref{eq:F_two_sided_selection}; this appendix proves the sender embedding and that identity. Write $\Sel_S\in\{0,1\}^{M_S\times M_S^2}$ for the source-side counterpart of $\Sel_R$, with $(\Sel_S)_{k,\,(c-1)M_S+d}=\delta_{k,c}\,\delta_{c,d}$.

Source incoherence restricts the source mutual intensity to its diagonal, $\avg{\vj_i\vj_i^{\dagger}}=\diagop(\vb{B})$ with $\vb{B}=\avg{\lvert\vj_i\rvert^{\circ 2}}$ (Sec.~\ref{sec:incoherent_MI}), and the source-side selection realizes this diagonal as the channel input,
\begin{equation}\label{eq:sender_embed}
\vecop\!\big(\diagop(\vb{B})\big)=\Sel_S^{\top}\vb{B}.
\end{equation}
Evaluating the right-hand side at the source flat index $(c-1)M_S+d$,
\[
\begin{aligned}
\big[\Sel_S^{\top}\vb{B}\big]_{(c-1)M_S+d}
&=\sum_{k}(\Sel_S)_{k,(c-1)M_S+d}\,B_k\\
&=\sum_{k}\delta_{k,c}\,\delta_{c,d}\,B_k
=\delta_{c,d}\,B_c,
\end{aligned}
\]
the $(c,d)$ entry of $\diagop(\vb{B})$; the two vectors agree in every component, which proves Eq.~\eqref{eq:sender_embed}.

The two-sided diagonal selection of the lifted propagation operator is the Hadamard square of the Green's function, an instance of the Hadamard product's realization as a selection of the Kronecker product~\cite{visick2000quantitative},

\begin{equation}\label{eq:hadamard}
\Sel_R\big(\Gt\otimes\Gt^{*}\big)\Sel_S^{\top}=\Gt^{*}\odot\Gt.
\end{equation}
Fix $p\in\{1,\dots,M_R\}$ and $k\in\{1,\dots,M_S\}$ and write the $(p,k)$ entry as a sum over the inner receiver pair $(a,b)$ and source pair $(c,d)$, inserting the selector entries,
\[
\begin{aligned}
&\big[\Sel_R(\Gt\otimes\Gt^{*})\Sel_S^{\top}\big]_{pk}\\
&\quad=\sum_{a,b}\sum_{c,d}
\delta_{p,a}\,\delta_{a,b}\,\delta_{k,c}\,\delta_{c,d}\,
\big(\Gt\otimes\Gt^{*}\big)_{\substack{(a-1)M_R+b,\\(c-1)M_S+d}}.
\end{aligned}
\]
The Kronecker deltas enforce $a=b=p$ and $c=d=k$, leaving the single entry at the diagonal receiver pair $(p,p)$ and diagonal source pair $(k,k)$. With the conjugate on the second factor, the lifted Kronecker product has entries $(\Gt\otimes\Gt^{*})_{(p-1)M_R+p',\,(k-1)M_S+l}=(\Gt)_{pk}\,(\Gt^{*})_{p'l}$; at $p'=p$, $l=k$,
\[
\begin{aligned}
&\big[\Sel_R(\Gt\otimes\Gt^{*})\Sel_S^{\top}\big]_{pk}\\
&\quad=\big(\Gt\otimes\Gt^{*}\big)_{(p-1)M_R+p,\,(k-1)M_S+k}\\
&\quad=(\Gt)_{pk}\,(\Gt^{*})_{pk}=\lvert(\Gt)_{pk}\rvert^{2}\\
&\quad=(\Gt^{*}\odot\Gt)_{pk}.
\end{aligned}
\]
The identity holds for every $(p,k)$, which proves Eq.~\eqref{eq:hadamard}.

Composing the lifted propagation of Eq.~\eqref{eq:kronecker_lifted_field}, the receiver selection of Eq.~\eqref{eq:lifted_mutual_field_sel}, source incoherence, and the sender embedding of Eq.~\eqref{eq:sender_embed} yields $\avg{\lvert\ve_R\rvert^{\circ 2}}=\Sel_R(\Gt\otimes\Gt^{*})\Sel_S^{\top}\vb{B}=(\Gt^{*}\odot\Gt)\vb{B}=\Ft\vb{B}$, the incoherent channel operator and channel law of the main text. The reduction to $\Ft$ follows from source incoherence rather than from square-law detection alone: incoherence nulls the off-diagonal source coherences $\avg{(\vj_i)_k(\vj_i)_l^{*}}$, $k\neq l$, the restriction that $\Sel_S^{\top}$ imposes at the operator level and the same cross-term structure whose survival makes an interferometric front end phase-bearing in the coherent channel of Eq.~\eqref{eq:cross_term_intensity}.

\subsection{Modal alignment and the output-entropy subadditivity converse}\label{asec:optimal_product_channel}
For a linear channel $\vb{H}=U\Sigma V^{\dagger}$ with isotropic additive noise $\vb{n}\sim\mathcal{CN}(\vb 0,N\I)$ and a Gaussian input of covariance $\vb{Q}=\EX[\vb{x}\vb{x}^{\dagger}]$, the mutual information is
\begin{equation}
\begin{aligned}
\mathcal{I}(\vb{x};\vb{y})&=\log\det\!\Big(\I+\frac{1}{N} \vb{H}\vb{Q}\vb{H}^{\dagger}\Big)\\
&=\log\det\!\Big(\I+\tfrac1N V^{\dagger}\vb{Q}V \Sigma^{2}\Big),
\end{aligned}
\label{opc:rate}
\end{equation}
obtained by invariance of the determinant under unitary similarity, and $\det(\I+AB)=\det(\I+BA)$. Over the real field the same expression carries an overall factor of one half. The mutual information depends on the channel only through the gains \(\Sigma^{2}\) and on the rotated covariance $V^{\dagger}\vb{Q}V$, the input covariance written in the channel's right-singular basis. For fixed eigenvalue spectra of $\Sigma^{2}$ and $\vb{Q}$, it is largest when $V^{\dagger}\vb{Q}V$ is diagonal with the eigenvalues of $\vb{Q}$ sorted against the gains, i.e. when the right-singular basis $V$ is aligned with the eigenbasis of $\vb{Q}$ and the largest eigenvalue pairs with the largest gain. The channel is then the scalar product $\tilde y_i=\sigma_i\tilde x_i+\tilde n_i$ in the coordinates $\tilde{\vb{x}}=V^{\dagger}\vb{x}$, $\tilde{\vb{y}}=U^{\dagger}\vb{y}$. Under these conditions, the entropies are as follows:
\begin{equation}
\begin{aligned}
h(\tilde{\vb{y}}\mid\tilde{\vb{x}})&=\sum_i h(\tilde y_i\mid\tilde x_i),\\
h(\tilde{\vb{y}})&\le\sum_i h(\tilde y_i),\\
\mathcal{I}(\vb{x};\vb{y})&\le\sum_i \mathcal{I}(\tilde x_i;\tilde y_i),
\end{aligned}
\label{ffpf:subadd}
\end{equation}
the inequality by subadditivity, with equality for independent inputs \cite{telatar1999capacity,tse2005fundamentals,cover2006elements}. Which matrix/operator is rotated to reach the alignment depends only on which is free: with the channel fixed one can precode the input so that the eigenvectors of \(\vb{Q}\) align with $V$; while with the input fixed one designs the right singular vectors of the channel to fit the eigenvectors of $\vb{Q}$ instead \cite{carson2012communications}. The left-singular basis $U$ does not enter Eq.~\eqref{opc:rate}, because isotropic (white) noise lets $U^{\dagger}$ be undone at the receiver without loss. This holds for coherent (field) detection; Appendix~\ref{asec:farfield_focusing} shows that intensity detection removes it and fixes $U$ as well.

For intensity detection, with the detected intensities $y_{k}=\lvert(\vb{H}\vb{x}+\vb{n})_{k}\rvert^{2}$ collected in $\vb{y}$, the subadditivity argument extends to a converse over front ends. 
In information theory, an achievability result exhibits one scheme attaining a given mutual information. A converse bounds every admissible scheme. Here the schemes are the front ends $\vb{H}$ at fixed input statistics~\cite{cover2006elements}.
Conditioned on the input, the noisy field components $(\vb{H}\vb{x}+\vb{n})_k$ are independent across $k$, because the noise is white. Each intensity $y_k$ is a function of its own component alone. The intensities are therefore conditionally independent,
\begin{equation}
\gamma_{\vb{y}\mid\vb{x}}=\prod_{k=1}^{M_R}\gamma_{y_k\mid\vb{x}},
\qquad
h(\vb{y}\mid\vb{x})=\sum_{k=1}^{M_R}h(y_k\mid\vb{x}).
\end{equation}
The unconditional output entropy obeys the independence bound $h(\vb{y})\le\sum_k h(y_k)$~\cite{cover2006elements}. Subtracting the conditional identity yields, for every front end $\vb{H}$ and every input distribution,
\begin{equation}
\mathcal{I}(\vb{x};\vb{y})=h(\vb{y})-h(\vb{y}\mid\vb{x})\le\sum_{k=1}^{M_R} \mathcal{I}(\vb{x};y_{k}).
\end{equation}
Each term in the bound is scalar. The intensity $y_k$ depends on the input only through the field component $(\vb{H}\vb{x})_k$, so $\mathcal{I}(\vb{x};y_k)=\mathcal{I}\big((\vb{H}\vb{x})_k;y_k\big)$. For a Gaussian input, $(\vb{H}\vb{x})_k\sim\mathcal{CN}\big(0,(\vb{H}\vb{Q}\vb{H}^{\dagger})_{kk}\big)$. Each term is therefore the scalar amplitude-detection mutual information of Appendix~\ref{asec:phase_retrieval_integral} at signal power $(\vb{H}\vb{Q}\vb{H}^{\dagger})_{kk}$. Equality holds if and only if $\vb{H}\vb{Q}\vb{H}^{\dagger}$ is diagonal. A diagonal $\vb{H}\vb{Q}\vb{H}^{\dagger}$ makes the jointly Gaussian pre-detection field components independent, hence the intensities independent, and the bound tight. A nonzero entry $(\vb{H}\vb{Q}\vb{H}^{\dagger})_{kl}$ gives $\operatorname{Cov}(y_{k},y_{l})=\lvert(\vb{H}\vb{Q}\vb{H}^{\dagger})_{kl}\rvert^{2}>0$~\cite{goodman2015statistical}, so the intensities are dependent and the bound is strict.

Each scalar term carries pre-log $\tfrac{1}{2}$~\cite{blachman1953comparison}, so the sum of $M_R$ terms bounds the pre-log by $M_R/2$, attained by the decoupled channel. Because the bound depends on the front end only through the diagonal entries $(\vb{H}\vb{Q}\vb{H}^{\dagger})_{kk}$, it establishes the optimality of decoupling only among front ends sharing those entries, and does not order front ends with different per-detector powers.

\section{The phase-retrieval channel -- coherent sources with square-law detection}\label{asec:app_phase_retrieval}
\subsection{Asymptotic expression for mode-sorted mutual information}\label{asec:phase_retrieval_integral}

\paragraph{Channel model.} The transmitted signal is a zero-mean proper complex Gaussian current vector $\vb{x}\sim\mathcal{CN}(0,\vb{Q})$, $\vb{x}\in\mathbb{C}^{M_S}$. Through the linear map representing the structured Green's-function channel \(\G_{t,RS}\in\mathbb{C}^{M_R\times M_S}\) (written as $\G$ for notational shorthand), the signal produces a noiseless propagated field $\ve_R=\G\vb{x}$ which undergoes additive white circular noise before square-law readout:
\begin{equation}
\tilde{\ve}_R:=\ve_R+\vn=\G\vb{x}+\vn,\quad \vn\sim\mathcal{CN}(0,N \I_{M_R}),\quad \vb{x}\perp \vn,
\end{equation}
with the detector recording the elementwise intensity $\vb{y}=|\tilde{\ve}_R|^{\circ2}$, i.e.\ $y_i=|(\tilde{\ve}_R)_i|^2$. Here $N>0$ is the complex-noise variance ($\mathbb{E}_{n_i}[|n_i|^2]=N$), so the real and imaginary parts of each $n_i$ are i.i.d.\ $\mathcal N(0,N/2)$. The total pre-detection covariance is
\begin{equation}
\vb{K}:=\mathbb{E}_{\vb{x},\vb{n}}[\tilde{\ve}_R \tilde{\ve}_R^{\dagger}]=\G \vb{Q}\G^{\dagger}+N \I_{M_R},
\end{equation}

\noindent with the cross terms vanishing since $\vb{x},\vb{n}$ are independent and zero-mean.

\paragraph{Amplitude vs. intensity.} We work with the amplitude variable $a_i:=|(\tilde{\ve}_R)_i|$ as output rather than the intensity $y_i=a_i^2$. Because $t\mapsto t^2$ is a bijection on \(\mathbb{R}_{\geq 0}\) and there is no post-detection noise in this model, the two carry identical information:
\begin{equation}
\label{eq:amp-int}
\mathcal{I}(\vb{x};\vb{a})=\mathcal{I}(\vb{x};\vb{a}^{\circ 2})=\mathcal{I}(\vb{x};\vb{y}).
\end{equation}
The figure of merit is the mutual information of Eq.~\eqref{eq:mutual_info_integral_universal}, specialized to the amplitude output:
\begin{equation}
\label{eq:MI}
\mathcal{I}(\vb{x};\vb{a})=\int_{\mathbb{C}^{M_S}}\int_{\mathbb{R}_{\geq 0}^{M_R}}
\gamma_{\vb x}(\vb{x})\gamma_{\vb a | \vb x}(\vb{a}|\vb{x})
\log\frac{\gamma_{\vb a|\vb x}(\vb{a} |\vb{x})}{\gamma_{\vb a}(\vb{a})}d\vb{a}d\vb{x},
\end{equation}
with $\gamma_{\vb a |\vb x}$, $\gamma_{\vb a}$ the conditional and marginal amplitude densities. The input current and the field are zero-mean proper complex Gaussians with covariances $\vb{Q}$ and $\vb{K}$,

\begin{equation}
\label{eq:gaussians}
\gamma_{\vb x}(\vb{x})=\frac{e^{-\vb{x}^{\dagger}\vb{Q}^{-1}\vb{x}}}{\pi^{M_S}\det \vb{Q}},
\qquad
\gamma_{\tilde{\ve}_R}(\tilde{\ve}_R)=\frac{e^{-\tilde{\ve}_R^{\dagger}\vb{K}^{-1}\tilde{\ve}_R}}{\pi^{M_R}\det \vb{K}}
\end{equation}
(if $\vb{K}$ is singular, we replace $\det \vb{K}$ with the pseudo-determinant and $\vb{K}^{-1}$ with the pseudoinverse $\vb{K}^{+}$). Amplitude detection and the field channel factor into the intermediate laws:
\begin{equation}
\label{eq:intermediate}
\begin{aligned}
\gamma_{\vb a |\tilde{\ve}_R}(\vb{a} |\tilde{\ve}_R)&=\prod_{i=1}^{M_R}\delta(a_i-|(\tilde{\ve}_R)_i|),\\
\gamma_{\tilde{\ve}_R |\vb x}(\tilde{\ve}_R |\vb{x})&=\frac{e^{-(\tilde{\ve}_R-\G\vb{x})^{\dagger}(\tilde{\ve}_R-\G\vb{x})/N}}{\pi^{M_R}N^{M_R}}.
\end{aligned}
\end{equation}
Combining the two conditional densities, $\gamma_{\vb a |\vb x}=\int_{\mathbb{C}^{M_R}}\gamma_{\vb a |\tilde{\ve}_R}\,\gamma_{\tilde{\ve}_R |\vb x}\,d\tilde{\ve}_R$, yields the explicit Rician product:
\begin{equation}
\label{eq:rician}
\gamma_{\vb a |\vb x}(\vb{a} |\vb{x})
=\prod_{i=1}^{M_R}\frac{2a_i}{N}\,e^{-(a_i^2+r_i^2)/N}\,I_0\!\Big(\frac{2a_i r_i}{N}\Big),
\end{equation}
where $r_i:=|\G_i\vb{x}|$ is the noiseless signal amplitude in mode $i$. This conditional law defines the channel.
The output density follows from either standard identity $\gamma_{\vb a}=\int\gamma_{\vb x}\gamma_{\vb a |\vb x}\,d\vb{x}=\int\gamma_{\tilde{\ve}_R}\gamma_{\vb a |\tilde{\ve}_R}\,d\tilde{\ve}_R$. Evaluating the latter, a proper complex Gaussian splits into amplitudes multiplied by an independent uniform phase:
\begin{equation}
\label{eq:marginal}
\gamma_{\vb a}(\vb{a})
=\frac{\prod_{i=1}^{M_R}a_i}{\pi^{M_R}\det \vb{K}}
\int_{[0,2\pi)^{M_R}} e^{-\tilde{\ve}_R^{\dagger}\vb{K}^{-1}\tilde{\ve}_R}\Big|_{|(\tilde{\ve}_R)_i|=a_i}\,d\boldsymbol\phi .
\end{equation}
We note that the integral over phases \(\boldsymbol{\phi}\) in \eqref{eq:marginal} is convex in $\vb{K}^{-1}$ (and, through the Woodbury matrix identity, in $\vb{Q}^{-1}$) by H\"older's inequality, but matrix inversion does not preserve curvature, thus $\gamma_{\vb a}$ is of indefinite curvature in the design variables $\vb{K},\vb{Q}$. A closed form exists when $\vb{K}$ is diagonal; that case is treated in Sec.~\ref{asec:PR_reduction}. The diagonal structure is also necessary for factorization: mutually independent output amplitudes imply uncorrelated output intensities, and the Gaussian fourth-moment identity gives $\operatorname{Cov}\big(|(\tilde{\ve}_R)_i|^2,|(\tilde{\ve}_R)_j|^2\big)=|\vb{K}_{ij}|^2$ for $i\neq j$, so independence forces every off-diagonal entry of $\vb{K}$ to vanish.

Both densities carry a fixed scale: $\gamma_{\vb a |\vb x}=N^{-M_R}g_2$ with $g_2(\vb{a} |\vb{x}):=N^{M_R}\gamma_{\vb a |\vb x}$, and $\gamma_{\vb a}=(\det \vb{K})^{-1}g_1$ with $g_1(\vb{a}):=\det \vb{K}\,\gamma_{\vb a}(\vb{a})$, the latter independent of the overall scale $\det \vb{K}$. Substituting into \eqref{eq:MI} and using $\int\gamma_{\vb x}\int\gamma_{\vb a |\vb x}\,d\vb{a}\,d\vb{x}=1$,

\begin{multline}
\label{eq:alpha}
\mathcal{I}(\vb{x};\vb{a})=\underbrace{\int_{\mathbb{C}^{M_S}}\int_{\mathbb{R}_{\geq 0}^{M_R}}\gamma_{\vb x}\,\gamma_{\vb a|\vb x}\,
\log\frac{g_2(\vb{a}|\vb{x})}{g_1(\vb{a})}\,d\vb{a}\,d\vb{x}}_{=:\ \alpha(\vb{Q})}\\
+\log\frac{\det \vb{K}}{N^{M_R}}.
\end{multline}
Since $\det \vb{K}=\det(N \I_{M_R}+\G \vb{Q}\G^{\dagger})=N^{M_R}\det(\I_{M_R}+\tfrac1N\G \vb{Q}\G^{\dagger})$, the constant term is the coherent log-determinant, and
\begin{equation}
\label{eq:alpha-final}
\mathcal{I}(\vb{x};\vb{a})=\alpha(\vb{Q})+\log\det\!\Big(\I_{M_R}+\tfrac1N\G \vb{Q}\G^{\dagger}\Big),
\quad \alpha(\vb{Q})\le 0,
\end{equation}
with the second term being the coherent (phase-sensitive) mutual information $\mathcal{I}(\vb{x};\tilde{\ve}_R)$, and $\alpha\le0$ the mutual information lost to envelope detection (nonpositive by the data-processing inequality $\vb{x}\to\tilde{\ve}_R\to\vb{a}$; we note $\alpha=0$ in the coherent case, where the observable is the field $\tilde{\ve}_R$ itself and the mutual information equals the log-determinant exactly). The remainder of this appendix evaluates $\alpha$, equivalently $\mathcal{I}$, in the high-SNR receiver-limited regime.

\subsubsection{Diagonal output covariance regime: reduction to scalar product channels} \label{asec:PR_reduction}
\noindent The only assumption used is that the effective pre-detection covariance is diagonal in the detector basis,
\begin{equation}
\label{eq:diag}
\G \vb{Q}\G^{\dagger}=\diagop(s_1,\dots,s_{M_R}),\quad s_i\ge0,
\end{equation}
so $\vb{K}=\diagop(k_1,\dots,k_{M_R})$ with $k_i=N+s_i$. We note that this does \emph{not} require $\G$ or $\vb{Q}$ to be diagonal in the physical basis; it states that the detector-mode fields are decorrelated, i.e.\ the rows of $\G$ are orthogonal under the $\vb{Q}$-weighted inner product, as realized by a modal front end that sorts modes \emph{before} square-law detection. The ordering is nontrivial: a unitary $\vb{U}$ applied after the square law cannot decouple the channel, since in general $\vb{U}^{\dagger}|\vb{U}\vb{v}+\vb{n}|^{\circ 2}\neq|\vb{v}+\vb{U}^{\dagger}\vb{n}|^{\circ 2}$ for a field $\vb{v}$; the diagonalizing modal operation must act on the field, prior to $|\cdot|^{\circ 2}$. Per-mode, we use the signal power $s_i=(\G \vb{Q}\G^{\dagger})_{ii}$, the average pre-detection SNR $\rho_i:=s_i/N$, and the output power $k_i=N(1+\rho_i)$. High SNR means $\rho_i\to\infty$; concretely, we set $\vb{Q}_t=t\vb{Q}_0$ and let $t\to\infty$, such that every mode with $(\G \vb{Q}_0\G^{\dagger})_{ii}>0$ has $\rho_i\to\infty$ at fixed $N$.

For diagonal $\vb{K}$ the phase integral in \eqref{eq:marginal} closes and the output marginal becomes a product of Rayleigh laws,

\begin{equation}
\label{eq:rayleigh}
\gamma_{\vb a}(\vb{a})=\prod_{i=1}^{M_R}\frac{2a_i}{k_i}\,e^{-a_i^2/k_i}.
\end{equation}
Under the change of variables $y_i=a_i^2$, the mode-$i$ Rician factor of Eq.~\eqref{eq:rician} is a noncentral chi-squared density with two degrees of freedom (noncentrality parameter $2r_i^2/N$, scale $N/2$), and the mode-$i$ factor of Eq.~\eqref{eq:rayleigh} is an exponential density with mean $k_i$; these are the intensity-domain forms cited in Sec.~\ref{sec:phase_retrieval}. The conditional \eqref{eq:rician} is already a product, due to the modal noises being independent. The ratio of the two products factorizes over $i$, and taking the logarithm in \eqref{eq:MI} turns the product into a sum, giving the exact decomposition:

\begin{equation}
\label{eq:sum}
\mathcal{I}(\vb{x};\vb{a})=\sum_{i=1}^{M_R} \mathcal{I}_i,\qquad
\mathcal{I}_i:=\mathbb{E}_{\vb{x}}\,\mathbb{E}_{n_i}\!\left[\log\frac{\gamma_i(a_i |\vb{x})}{\gamma_i(a_i)}\right],
\end{equation}

where $\gamma_i(a_i |\vb{x})=\frac{2a_i}{N}e^{-(a_i^2+r_i^2)/N}I_0(2a_i r_i/N)$ and $\gamma_i(a_i)=\frac{2a_i}{k_i}e^{-a_i^2/k_i}$ are the mode-$i$ factors of \eqref{eq:rician} and \eqref{eq:rayleigh}, and the inner expectation is over the mode-$i$ noise $n_i$ with $\vb{x}$ held fixed. Because $\G_i\vb{x}\sim\mathcal{CN}(0,s_i)$, the noiseless modal intensity $r_i^2=|\G_i\vb{x}|^2$ is exponential with mean $s_i$, so $r_i^2/s_i$ is unit-exponential; likewise $a_i^2/k_i$ is unit-exponential since $(\tilde{\ve}_R)_i\sim\mathcal{CN}(0,k_i)$. For a unit-exponential $E$,
\begin{equation}
\label{eq:expmoments}
\mathbb{E}[\log E]=-\gamma_E,\qquad \mathbb{E}[(\log E)^2]=\gamma_E^2+\tfrac{\pi^2}{6},
\end{equation}
so in particular $\mathbb{E}_{\vb{x}}\mathbb{E}_{n_i}[\log a_i^2]=\log k_i-\gamma_E$.

Written in terms of differential entropies, the mode-$i$ mutual information of Eq.~\eqref{eq:sum} is $\mathcal{I}_i=h(a_i)-h(a_i|\vb{x})$, where $h(a_i):=-\mathbb{E}_{\vb{x}}\mathbb{E}_{n_i}[\log\gamma_i(a_i)]$ and $h(a_i|\vb{x}):=-\mathbb{E}_{\vb{x}}\mathbb{E}_{n_i}[\log\gamma_i(a_i|\vb{x})]$ are the marginal and conditional differential entropies of the mode-$i$ amplitude~\cite{cover2006elements}.

The marginal entropy is exact. From the Rayleigh factor of Eq.~\eqref{eq:rayleigh}, $-\log\gamma_i(a_i)=\log\tfrac{k_i}{2}-\tfrac12\log a_i^2+\tfrac{a_i^2}{k_i}$; taking the expectation with $\mathbb{E}_{\vb{x}}\mathbb{E}_{n_i}[a_i^2]=k_i$ and the logarithmic moment above,
\begin{equation}\label{eq:PR_hmarg}
h(a_i)=\tfrac12\log k_i+1-\log2+\tfrac{\gamma_E}{2}.
\end{equation}

The conditional entropy tends to the entropy of the in-phase noise quadrature. Condition on $\vb{x}$ and rotate the mode-$i$ complex plane by $e^{-i\arg(\G_i\vb{x})}$; circular noise is invariant under the rotation, so in distribution $a_i=|r_i+n_\parallel+i\,n_\perp|$ with $n_\parallel,n_\perp$ i.i.d.\ real $\mathcal N(0,N/2)$, and the conditional law depends on $\vb{x}$ only through $r_i$; write $h(a_i|r_i)$ for its differential entropy at fixed $r_i$. For $r_i^2\gg N$ the envelope expands as $a_i=r_i+n_\parallel+n_\perp^2/(2r_i)+O(|n_i|^3/r_i^2)$: the randomness of $a_i$ is the single Gaussian quadrature $n_\parallel$, up to corrections of relative size $\sqrt{N}/r_i$, and differential entropy is unchanged by the shift $r_i$, so $h(a_i|r_i)\to\tfrac12\log(\pi e N)$, the entropy of $\mathcal N(0,N/2)$, as $r_i^2/N\to\infty$~\cite{rice1945mathematical,goodman2015statistical}. Moreover $h(a_i|r_i)$ is continuous in $r_i/\sqrt{N}$ with finite limits at both ends (the Rayleigh value $\tfrac12\log N+1-\log2+\tfrac{\gamma_E}{2}$ at $r_i=0$ and the Gaussian value at infinity), and therefore bounded. Since $r_i^2/N=\rho_i\,(r_i^2/s_i)$ with $r_i^2/s_i$ unit-exponential, every fixed realization of the unit-exponential factor has $r_i^2/N\to\infty$ as $\rho_i\to\infty$, and bounded convergence gives
\begin{equation}\label{eq:PR_hcond}
h(a_i|\vb{x})=\mathbb{E}_{\vb{x}}\big[h(a_i|r_i)\big]=\tfrac12\log(\pi e N)+o(1).
\end{equation}

Subtracting Eq.~\eqref{eq:PR_hcond} from Eq.~\eqref{eq:PR_hmarg}, with $k_i/N=1+\rho_i$ and $\log2+\tfrac12\log(\pi e)=\tfrac12+\tfrac12\log(4\pi)$, each active mode obeys the scalar half-log law
\begin{equation}\label{eq:PR_scalar}
\mathcal{I}_i=\tfrac12\log(1+\rho_i)+c_0+o(1),\quad c_0=\tfrac12+\tfrac{\gamma_E}{2}-\tfrac12\log(4\pi),
\end{equation}
with $c_0\approx-0.4769$ nats. Modes with $s_i=0$ satisfy $\gamma_i(a_i|\vb{x})=\gamma_i(a_i)$ and contribute zero both to Eq.~\eqref{eq:sum} and to the log-determinant. Summing Eq.~\eqref{eq:PR_scalar} over the modes, using Eq.~\eqref{eq:diag} to identify $\sum_i\log(1+\rho_i)=\log\det\big(\I_{M_R}+\tfrac{1}{N}\G\vb{Q}\G^{\dagger}\big)$, and passing from $\vb{a}$ to $\vb{y}$ through Eq.~\eqref{eq:amp-int}, we arrive at the half-log-det law: for $\rho_i\to\infty$ in every mode,
\begin{equation}\label{eq:PR_half_logdet}
\begin{aligned}
\mathcal{I}(\vb{x};\vb{y})&=\tfrac12\,\log\det\!\Big(\I_{M_R}+\tfrac1N\G\vb{Q}\G^{\dagger}\Big)+M_R c_0+o(1),\\
c_0&=\tfrac12+\tfrac{\gamma_E}{2}-\tfrac12\log(4\pi).
\end{aligned}
\end{equation}
If only $r<M_R$ diagonal entries of $\G\vb{Q}\G^{\dagger}$ are positive, the inactive modes contribute zero on both sides and $M_R c_0$ is replaced by $r c_0$. Under the scaling $\vb{Q}_t=t\vb{Q}_0$ the leading term grows as $\tfrac{M_R}{2}\log t$: one real radial degree of freedom per detector mode, at half the coherent pre-log. Comparing with Eq.~\eqref{eq:alpha-final} identifies the high-SNR value of the envelope-detection loss, $\alpha(\vb{Q})=-\tfrac12\log\det\big(\I_{M_R}+\tfrac{1}{N}\G\vb{Q}\G^{\dagger}\big)+M_R c_0+o(1)$: square-law detection removes exactly half of the leading coherent log-determinant growth.

\subsection{Derivation of surrogate objective for photonic optimization of phase retrieval}\label{asec:surrogate_objective}

The exact mutual information of the phase-retrieval channel, Eq.~\eqref{eq:mutual_information_integral_alpha}, has no closed form for general \(\G_{t,RS}\), \(\vb{Q}\), or \(N\), and the Monte Carlo estimator of Appendix~\ref{asec:numerics_phase_retrieval} is costly and noisy to differentiate through. For the gradient-based inverse design of the front end we therefore optimize a surrogate objective that holds the output intensity and the additive noise as approximately Gaussian in \(\mathbb{R}^{M_R}_{\geq 0}\), with covariances matched to their exact distributions, and reserve the Monte Carlo integral for evaluation of the reported mutual information.

Leveraging the lifted representation of Eq.~\eqref{eq:lifted_mutual_field_sel} to rewrite the channel model of Eq.~\eqref{eq:coh_to_inc_channel_law},
\begin{subequations}
\begin{align}
&\vb{y} = \Sel_R\big(\G_{t,RS} \otimes \G_{t,RS}^{*}\big) \operatorname{vec}\!\big(\vb{x}\vb{x}^{\dagger}\big) + N\mathbf{1} + \vb{w}, \\ &\vb{w}_k := \big|(\vb{e}_R + \vb{n})_k\big|^2 - s_k - N,
\end{align}
\end{subequations}
with \(N\mathbf{1} = \mathbb{E}_{\vb{n}}[\,|\vb{n}|^{\circ 2}]\) the readout floor, \(s_k := |(\vb{e}_R)_k|^2\) the noiseless modal intensity, and \(\vb{w}\) zero-mean. The two second moments the surrogate requires are available in closed form: the Gaussian fourth-moment identity gives the signal covariance \(\operatorname{Cov}(\vb{s}) = \Gamma \odot \Gamma^*\) with \(\Gamma := \G_{t,RS}\vb{Q}\G_{t,RS}^{\dagger}\), and freezing the conditional noise covariance over the signal gives \(\Sigma_{\vb{w}} = 2N\,D(\Gamma) + N^2 \Id_{M_R}\), \(D(\Gamma) := \Id_{M_R}\odot\Gamma\). Replacing the rank-one lifted signal and the Rician post-detection noise by Gaussians of matched covariance yields a closed-form, everywhere-differentiable surrogate for the phase-retrieval mutual information,
\begin{multline} \label{eq:surrogate_phase_retrieval_objective}
    \mathcal{I}\big(\vb{x}; \vb{y}\big) \approx \mathcal{I}_{\mathrm{PR}}(\G_{t,RS}) \\
    = \tfrac{1}{2}\log\det\!\Big(\Id_{M_R} + \big[\,2N\,D(\Gamma) + N^2\Id_{M_R}\,\big]^{-1}\big(\Gamma \odot \Gamma^* \big)\Big).
\end{multline}
Because neither \(\big|\vb{e}_R\big|^{\circ 2}\) nor \(\vb{w}\) is Gaussian-distributed---a Gaussian signal of matched covariance can only raise the mutual information, while Gaussian noise of matched covariance can only lower it---\(\mathcal{I}_{\mathrm{PR}}\) is neither an upper nor a lower bound for the exact mutual information \(\mathcal{I}\big(\vb{x}; \vb{y}\big) = \mathcal{I}\big(\vb{x}\vb{x}^{\dagger}; \vb{y}\big)\). For \(M_R = M_S\) and \(\Gamma\) diagonal by mode-focusing, in the high-SNR limit Eq.~\eqref{eq:surrogate_phase_retrieval_objective} reproduces the pre-log and leading half-log-det term of Eq.~\eqref{eq:half_logdet_asymptotic_diagonal_hiSNR}, with per-mode constant \(-\tfrac12\log 2\) in place of \(c_0\).

The surrogate defines the structural optimization used to shape the front ends of Fig.~\ref{fig:interferometric_sources}:
\begin{subequations} \label{eq:surrogate_PR_topopt}
\begin{align}
\max_{\varepsilon(\vb{r}_D)} \quad & \mathcal{I}_{\mathrm{PR}}\big(\G_{t,RS}(\varepsilon)\big) = \\
\max_{\varepsilon(\vb{r}_D)} &\tfrac12\log\det\!\Big(\Id_{M_R} + \big[\,2N\,D(\Gamma) + N^2\Id_{M_R}\,\big]^{-1} \nonumber\\
&\hspace{6em}\times\big(\Gamma\odot\Gamma^*\big)\Big) \\
\text{s.t.}\quad & \Gamma = \G_{t,RS}(\varepsilon)\,\vb{Q}\,\G_{t,RS}^{\dagger}(\varepsilon), \\
& \big(\nabla\times\nabla\times \,-\, \varepsilon(\vb{r})\,\omega^2\big)\,\G_t(\vb{r},\vb{r}';\varepsilon) = \omega^2\,\vb{I}\,\delta(\vb{r}-\vb{r}').
\end{align}
\end{subequations}
The Maxwell constraint restricts \(\G_{t,RS}(\varepsilon)\) to realizable structures, so the optimization is over physically admissible front ends rather than arbitrary operators; the design \(\varepsilon\) shapes the channel while \(\vb{Q}\) allocates power across the source modes, the two levers of Sec.~\ref{sec:intro}. In the numerical results \(\vb{Q}\) is held fixed at the isotropic \((P/M_S)\Id_{M_S}\) with power budget \(P\), and the optimization ranges over \(\varepsilon\) alone.

Because \(\mathcal{I}_{\mathrm{PR}}\) is a closed-form function of \(\Gamma = \G_{t,RS}\vb{Q}\G_{t,RS}^{\dagger}\), its gradient with respect to the channel \(\partial\mathcal{I}_{\mathrm{PR}}/\partial\G_{t,RS}\) is available analytically~\cite{palomar2006gradient}, or equivalently by reverse-mode differentiation of the log-determinant. The structural gradient \(\partial\mathcal{I}_{\mathrm{PR}}/\partial\varepsilon(\vb{r}_D)\) follows by composing the channel gradient with the Maxwell adjoint \(\partial\G_{t,RS}/\partial\varepsilon\), exactly as detailed for the Monte Carlo estimator in Appendix~\ref{asec:numerics_phase_retrieval}. This construction requires one forward and one adjoint full-wave solve per source point and is independent of the number of detectors \(M_R\).

\subsection{Numerical Evaluation for Phase Retrieval}
\label{asec:numerics_phase_retrieval}

To enable gradient-based photonic inverse design, the mutual information of the phase-retrieval channel must be both numerically tractable and continuously differentiable with respect to the structural permittivity $\varepsilon$. We evaluate the mutual information using a nested, variance-reduced Monte Carlo (MC) estimator.

As derived in Eq.~\eqref{eq:alpha-final}, the mutual information is decomposed into the exact analytical coherent log-determinant and an envelope-detection information loss $\alpha \le 0$:
\begin{equation}
    \mathcal{I}(\vb{x}; \vb{y}) = \log \det\left(\I_{M_R} + \frac{1}{N} \G \vb{Q} \G^\dagger\right) + \alpha(\vb{Q}).
\end{equation}
Isolating the coherent log-determinant acts as a control variate, confining the Monte Carlo integration strictly to the information penalty induced by phase-blind detection. The exact penalty $\alpha(\vb{Q})$ is the expectation of the residual log-ratio:
\begin{equation}
    \alpha(\vb{Q}) = \mathbb{E}_{\vb{x}, \vn}\left[ \log \gamma_{\vb{a}|\vb{x}}(\vb{a}|\vb{x}) - \log \gamma_{\vb{a}}(\vb{a}) - c(\vb{x}, \tilde{\ve}_R) \right],
\end{equation}
where $c(\vb{x}, \tilde{\ve}_R) = \log \gamma_{\tilde{\ve}_R|\vb{x}}(\tilde{\ve}_R|\vb{x}) - \log \gamma_{\tilde{\ve}_R}(\tilde{\ve}_R)$ is the per-sample coherent log-ratio, whose exact expectation is the coherent log-determinant of Eq.~\eqref{eq:alpha-final}.

Because the marginal density $\gamma_{\vb{a}}(\vb{a})$ lacks a general closed-form expression for non-diagonal pre-detection covariances, we define an estimator $\hat{\alpha}(\vb{Q})$ using an outer average over $M_{\mathrm{MC}}$ samples and a nested inner Monte Carlo loop over $L_{\mathrm{MC}}$ samples drawn from the source prior, $\vb{x}' \sim \mathcal{CN}(0, \vb{Q})$:
\begin{equation}
\begin{aligned}
    \hat{\alpha}(\vb{Q}) = \frac{1}{M_{\mathrm{MC}}} \sum_{m=1}^{M_{\mathrm{MC}}} \bigg[ &\log \gamma_{\vb{a}|\vb{x}}(\vb{a}_m|\vb{x}_m)\\
    &- \log \Big( \tfrac{1}{L_{\mathrm{MC}}} \sum_{l=1}^{L_{\mathrm{MC}}} \gamma_{\vb{a}|\vb{x}}(\vb{a}_m | \vb{x}'_l) \Big)\\
    &- c(\vb{x}_m, \tilde{\ve}_{R,m}) \bigg].
\end{aligned}
\end{equation}
This nested estimator is biased. 
The inner sample mean is an unbiased estimator of $\gamma_{\vb{a}}(\vb{a})$, but because it sits inside a strictly concave logarithm, Jensen's inequality dictates that $\mathbb{E}_{\vb{x}'}[\log \hat{\gamma}_{\vb{a}}] \le \log \gamma_{\vb{a}}$. Consequently, the log-ratio is biased high, and $\hat{\alpha}$ exhibits a systematic positive $+O(1/L_{\mathrm{MC}})$ bias. The estimator therefore over-estimates the true mutual information, converging to the exact mutual information only for sufficiently large $L_{\mathrm{MC}}$.

\section{The incoherent imaging channel -- incoherent sources with square-law detection}\label{asec:app_incoherent}
\subsection{MI-optimality of focusing in the radiative far-field}\label{asec:farfield_focusing}

We determine the passive photonic front end that maximizes the mutual information of a spatially incoherent source radiating across a vacuum gap to an intensity detector, and show that, for a source occupying the diffraction-limited lattice of the geometry, intensity detection forces the front end to focus the field to points at every signal-to-noise ratio. The source occupies $S=[-D_S/2,D_S/2]$ and the detector $R=[-D_R/2,D_R/2]$, separated by a vacuum gap of length $L$, with $k_0=2\pi/\lambda=\omega/c$ and $\langle\,\cdot\,\rangle$ the temporal average defining the field correlations comprising the coherence matrix/mutual intensity.

A general-purpose imager operates without a scene model; the input distribution therefore cannot be fitted to a specific object ensemble: the least informative input consistent with a mean-power budget and the per-source variance constraint of the main text is, by maximum entropy, the Gaussian intensity $\vb B\sim\mathcal N(\boldsymbol{\mu}_B,\vb{Q}_B)$ with, from a flat per-source variance bound, the isotropic covariance $\vb{Q}_B=q\,\I$, $q>0$. The fixed mean $\boldsymbol{\mu}_B$ does not carry information, and together with the per-source $\kappa$-margin of the main text, holds the intensity nonnegative; the signal is the fluctuation $\vb B-\boldsymbol{\mu}_B$, and the channel is the linear Gaussian map
\begin{equation}
\vb y=\langle|\vb e_R|^{\circ 2}\rangle+\vb n=\F\,\vb B+\vb n,\qquad \vb n\sim\mathcal N(\vb 0,N\I),
\label{ffpf:linchan}
\end{equation}
with additive white detector noise $\vb n$ and Gaussian input $\vb B$ of covariance $\vb{Q}_B$. Writing the intensity operator $\F=U_{\F}\Sigma_{\F}V_{\F}^{\dagger}$, the mutual information is
\begin{equation}
\mathcal{I}=\tfrac12\log\det\!\Big(\I+\tfrac1N\,\F \vb{Q}_B\F^{\dagger}\Big)=\tfrac12\log\det\!\Big(\I+\tfrac1N\,V_{\F}^{\dagger}\vb{Q}_B V_{\F}\,\Sigma_{\F}^{2}\Big),
\label{ffpf:rate}
\end{equation}
the second equality following from unitary invariance of the determinant and the identity $\det(\I+AB)=\det(\I+BA)$ \cite{telatar1999capacity}. With the isotropic prior the conjugated covariance $V_{\F}^{\dagger}\vb{Q}_B V_{\F}=q\,\I$ is unchanged, and Eq.~\eqref{ffpf:rate} reduces to
\begin{equation}
\mathcal{I}=\tfrac12\sum_{n}\log\!\Big(1+\tfrac{q}{N}\,\sigma_{\F,n}^{2}\Big),
\label{ffpf:rateiso}
\end{equation}
a function of the gains $\Sigma_{\F}^{2}$ alone: an isotropic source does not distinguish a preferred basis, neither singular basis of the channel enters the mutual information, and the design is fixed entirely by the singular spectrum of $\F$. Eq.~\eqref{ffpf:rateiso} is increasing and Schur-concave in the gains $\sigma_{\F,n}^{2}$~\cite{marshall2011inequalities}; among realizable spectra of fixed total gain, it is therefore largest when the nonzero gains are as numerous and as equal as possible. With the input covariance fixed by the absence of a scene prior, the remaining freedom is the channel, and the optimization is over the front end that shapes $\F$ \cite{carson2012communications}.

The detector field produced by source current $j$ is the diffraction integral $e_R(x_R)=\frac{i}{k_0}\int_S\G_{0,L}(x_R,x_S)\,j(x_S)\,dx_S$, carried by the scalar vacuum (unstructured) Green's function; the propagation scalar $i/k_0$ is kept explicit throughout this appendix section, and $\G_{0,L}$ and the amplitude values $A_L$ below are the $\omega^2\mI\delta$-normalized free-space forms of the main text (Sec.~\ref{sec:channel_laws}). In the paraxial regime, valid when the transverse separations are small against the gap, $|x_R-x_S|\ll L$, the path length expands to second order and the propagator factors into a constant amplitude, two quadratic chirps, and a Fourier kernel,
\begin{equation}
\G_{0,L}(x_R,x_S)=A_L\,e^{ik_0x_R^2/2L}\,e^{-ik_0x_Rx_S/L}\,e^{ik_0x_S^2/2L},
\label{ffpf:prop}
\end{equation}
where the amplitude prefactor takes one of several geometry-dependent values, two of which are relevant here: $A_L=k_0^{2}\,e^{i(k_0L+\pi/4)}\,(8\pi k_0L)^{-1/2}$ for two-dimensional TM propagation between line apertures, the geometry of the main-text simulations, and $A_L=k_0^{2}\,e^{ik_0L}/(4\pi L)$ for two planar apertures across a three-dimensional gap, whose Fourier kernel is $e^{-ik_0\boldsymbol{\rho}_R\cdot\boldsymbol{\rho}_S/L}$ in two transverse dimensions; the results below depend on $A_L$ only through $|A_L|$ and hold for either geometry. The Fraunhofer approximation $L\gg D^{2}/\lambda$ is not made, as it would discard the quadratic phase across the apertures and is incompatible with the large-aperture regime $N_0=D_SD_R/\lambda L\gg1$ used below \cite{goodman2017introduction}. The two chirps are diagonal pure-phase operators that do not affect the singular values; the detector chirp is in addition erased by the squared modulus under direct readout and, under the front end below, cancelled by the conjugate quadratic phase $e^{-ik_0x_R^2/2L}$ that a thin lens of focal length $L$ can supply. Only the Fourier kernel remains along a given transverse axis $\hat{x}$,
\begin{equation}
\G_{0,L}(x_R,x_S)=A_L\,e^{-ik_0x_Rx_S/L},
\label{ffpf:fourier}
\end{equation}
the continuous, paraxial counterpart of the vacuum source-to-detector block $\G_{0,L}=\G_{0,RS}$ of the main text. Its singular system $\G_{0,L}=U_{\G}\Sigma_{\G}V_{\G}^{\dagger}$ is the prolate spheroidal system of the Slepian concentration problem, posed over functions whose spatial-frequency content is confined to the band $[-\Omega,\Omega]$ collected by the detector aperture,
\begin{equation}
\Omega=\frac{k_0D_R}{2L}=\frac{\pi D_R}{\lambda L}.
\label{ffpf:omega}
\end{equation}
The problem selects the functions most concentrated within the source window; the singular functions \(U_{\G}, V_{\G}\) are the prolate spheroidal functions and \(\lambda_n\) the concentration eigenvalues \cite{slepian1961prolate,landau1961prolate,slepian1976bandwidth}. These eigenvalues step from near unity to vanishing at the index
\begin{equation}
N_0=\frac{\Omega D_S}{\pi}=\frac{D_SD_R}{\lambda L},
\label{ffpf:n0}
\end{equation}
with a transition region of width $O(\log N_0)$ \cite{slepian1976bandwidth,franceschetti2015landau}; the singular values of $\G_{0,L}$ are therefore flat across the first $N_0$ modes and negligible beyond. $N_0$ is the number of orthogonal field modes the geometry supports; a passive element cannot increase the count. This can be seen in the trace of the concentration kernel, with a constant diagonal $\Omega/\pi$ integrating over $S$ to $\Omega D_S/\pi=N_0$. The resolvable cell is $\Delta x=\pi/\Omega=\lambda L/D_R$, the diffraction limit, here a consequence of where the singular spectrum collapses rather than of the first zero of a point-spread function \cite{franceschetti2017wave}, set entirely by the receiver aperture and independent of the source.

The count grows with $D_R$ only within the paraxial regime. A transverse spatial frequency propagates across the gap only if $|k_x|\le k_0$. The collected band therefore saturates at $\Omega=k_0\,\mathrm{NA}\le k_0$; the resolution saturates at the Abbe limit, $\Delta x\ge\lambda/2$; the mode count saturates at $N_0\le2D_S/\lambda$. In two transverse dimensions, the same construction gives $N_0=A_SA_R/(\lambda L)^{2}=A_S\,\Omega_{\mathrm{det}}/\lambda^{2}$, where $\Omega_{\mathrm{det}}=A_R/L^{2}$ is the detector solid angle \cite{toraldo1969degrees}. The small-angle form is the normal-incidence approximation of an exact phase-space count, and extrapolated to a full hemisphere it would give $2\pi A_S/\lambda^{2}$. The exact measure of propagating modes is $d^{2}k_{\perp}=k_0^{2}\cos\theta\,d\Omega$, filling a disk of area $\pi k_0^{2}$, so a detector collecting every propagating direction receives $N=A_S\,\pi k_0^{2}/(2\pi)^{2}=\pi A_S/\lambda^{2}$ modes, below the cut-set bound $A_S/(\lambda/2)^{2}=4A_S/\lambda^{2}$~\cite{franceschetti2017wave}.

We define an ideal passive front end $W$ acting on the fields immediately before the detectors. The front end is an abstraction: whichever structure implements it must realize the unitary image-plane Fourier transform $(We_R)(x)=(\lambda L)^{-1/2}\int_{-D_R/2}^{D_R/2}e^{ik_0 x_R x /L}\, e_R(x_R) \,dx_R$, with $x$ the image-plane coordinate; a lens performing the optical Fourier transform between conjugate planes is one realization. The map inverts the finite Fourier transform $\G_{0,L}$ performs and produces the tightest field spot the collected propagating band permits. The structured propagator $\G^{\star}_L=W\G_{0,L}$ has kernel
\begin{equation}
\begin{aligned}
\G^{\star}_L(x,x_S)&=\frac{A_L}{\sqrt{\lambda L}}\int_{-D_R/2}^{D_R/2}e^{ik_0x_R(x-x_S)/L}\,dx_R\\
&=\sqrt{\tfrac{\Omega}{\pi}}\,\sigma\,\sinc\!\big(\Omega(x-x_S)\big),\qquad \sigma=|A_L|\sqrt{D_R},
\end{aligned}
\label{ffpf:gstar}
\end{equation}
the diffraction-limited point-spread function: a source point at $x_S$ is imaged to a spot of width $\Delta x$ centered at $x_S$. Because $W$ is unitary, the gains and the count $N_0$ are unchanged,
\begin{equation}
(\G^{\star}_L)^{\dagger}\G^{\star}_L= \G_{0,L}^{\dagger}W^{\dagger}W\G_{0,L} =
\G_{0,L}^{\dagger}\G_{0,L}.
\label{ffpf:graminv}
\end{equation}
The spacing $\Delta x=\pi/\Omega$ is not arbitrary: the sinc point responses of Eq.~\eqref{ffpf:gstar} are mutually orthogonal as fields exactly at that spacing, and the zeros of $\sinc$ coincide with the zeros of $\sinc^{2}$, a property the intensity readout inherits below as the lattice diagonality $\sinc^{2}(\pi(m-n))=\delta_{mn}$. On the source-plane diffraction lattice $x_{S,m}=m\,\Delta x$ the captured fields are orthogonal: the detector-aperture overlap of the fields of two source cells is
\begin{equation}
\begin{aligned}
&\int_{-D_R/2}^{D_R/2}\G_{0,L}(x_R,x_{S,m})^*\,\G_{0,L}(x_R,x_{S,n})\,dx_R\\
&\quad=\sigma^{2}\,\sinc\!\big(\Omega(x_{S,m}-x_{S,n})\big)\\
&\quad=\sigma^{2}\,\sinc(\pi(m-n))=\sigma^{2}\,\delta_{mn},
\end{aligned}
\label{ffpf:ortho}
\end{equation}
unchanged by the unitary front end through Eq.~\eqref{ffpf:graminv}, the off-diagonals vanishing through cancellation of the signed oscillating tails; $W$ therefore carries the $N_0$ source cells to $N_0$ mutually orthogonal detector spots at a single gain $\sigma$. In the localized source-cell basis $V_{\mathrm D}$ and detector-spot basis $U_{\mathrm D}$, the subscript denoting the design basis, the structured propagator is therefore diagonal,
\begin{equation}
\G^{\star}_L=U_{\mathrm D}\,\Sigma_{\G}\,V_{\mathrm D}^{\dagger}.
\label{ffpf:gstardiag}
\end{equation}
A degenerate singular value fixes the singular bases only up to a common unitary rotation; the flat spectrum, achieved via focusing, therefore leaves the front end free to choose the localized bases $U_{\mathrm D},V_{\mathrm D}$ over the prolate bases $U_{\G},V_{\G}$. Equation~\eqref{ffpf:gstardiag} marks the change of working coordinates: the continuum description in prolate wave functions on the source and receiver planes is set aside, statements from here on are indexed on the lattice, $n$ labeling the source cell at $x_{S,n}$ and $m$ the point detector at the image coordinate $x_{S,m}$, the two descriptions connected by a unitary rotation within the degenerate plateau subspace.

The channel Eq.~\eqref{ffpf:linchan} is carried by the Hadamard intensity operator $\F^{\star}_L=k_0^{-2}\,\G^{\star}_L\odot\G^{\star,*}_L$, the far-field realization of the main-text PSF channel $\F_{t,RS}=\G_{t,RS}^{*}\odot\G_{t,RS}$ (the main text absorbs the propagation scalar into $\G_{t,RS}$; here the scalar stays explicit and enters the intensity map as the fixed factor $k_0^{-2}$), with kernel $\F^{\star}_L(x_R,x_S)=k_0^{-2}\,|\G^{\star}_L(x_R,x_S)|^{2}$ in continuous variables. By Eq.~\eqref{ffpf:rateiso}, the mutual information is set by the number of near-equal singular values of $\F$, maximized by a localized front end. Under the squared modulus, a field mode confined to one spot squares to a single localized intensity and contributes one independent mode of $\F$. A field mode spread across several cells squares to an intensity overlapping its neighbors and reduces the rank.

The unstructured vacuum propagator transfers zero spatial structure to the intensity operator. Because $|\G_{0,L}|=|A_L|$ is constant, $\F_{0,L}=k_0^{-2} |A_L|^{2}$ has rank one, and the detector measures the total source power alone. The spatial structure of the source survives in the two-point coherence of the detector field, which the van Cittert--Zernike relation identifies as the Fourier transform of the source intensity \cite{goodman2015statistical}. Point-wise intensity detection retains only the uniform diagonal of that coherence; the front end below converts the off-diagonal coherence into detector-plane intensity structure.

The focused operator, by contrast, inherits the full count of nontrivial modes,
\begin{equation}
\F^{\star}_L(x_R,x_S)=k_0^{-2}\,|\G^{\star}_L(x_R,x_S)|^{2}\propto\sinc^{2}\!\big(\Omega(x_R-x_S)\big),
\label{ffpf:Ffoc}
\end{equation}
recovering the diffraction-limited intensity point-spread function, whose Fourier transform is the autocorrelation of the aperture \cite{toraldo1969degrees}. On the diffraction lattice, $\sinc^{2}(\pi(m-n))=\delta_{mn}$, and $\F^{\star}_L$ restricted to the lattice cells acts as $\sigma^{2}/k_0^{2}$ times the identity, each point detector collecting one source cell. 
Focusing therefore transfers the entire Slepian budget of Eq.~\eqref{ffpf:n0} into the intensity operator as a flat spectrum on all $N_0$ modes. 

The leading singular value of a Hadamard product is bounded by the product of the leading singular values of its factors \cite{ando1987singular,horn1991topics}, hence $\sigma_{\F,1}\le\sigma^{2}/k_0^{2}$ for every passive front end, the field channel gains being invariant under a unitary $W$ by Eq.~\eqref{ffpf:graminv}; Eq.~\eqref{ffpf:rateiso} increases in each gain, a lattice source of $N_0$ cells admits at most $N_0$ nonzero gains, and focusing holds all $N_0$ at the bound $\sigma^{2}/k_0^{2}$; the focused spectrum dominates every realizable spectrum gain by gain. 

The flat spectrum of Eq.~\eqref{ffpf:Ffoc} leaves the mutual information Eq.~\eqref{ffpf:rateiso} invariant under any orthogonal rotation of the modes. Every entry of the intensity operator is a squared field amplitude, so $\F\ge0$ entrywise. A nonnegative operator with a flat singular spectrum is a scaled permutation, and the orthogonal freedom reduces to a relabeling of which source cell is carried to which detector spot; any superposition mixing distinct spots would require the signed cancellations a nonnegative operator cannot produce. By Eq.~\eqref{ffpf:gstar} the structured field at the image point $x_{S,m}$ is the sinc-weighted integral $(i/k_0)\sqrt{\Omega/\pi}\,\sigma\int_S\sinc\!\big(\Omega(x_{S,m}-x_S)\big)\,j(x_S)\,dx_S$; squaring and averaging over the incoherent phases yields:
\begin{equation}
\begin{aligned}
y_m&=\big\langle|(i/k_0)(\G^{\star}_L j)(x_{S,m})|^{2}\big\rangle+\eta_m=\frac{\sigma^{2}}{k_0^{2}}B_m+\eta_m,\\
B_m&=\frac{\Omega}{\pi}\int_S\sinc^{2}\!\big(\Omega(x_{S,m}-x_S)\big)\,B(x_S)\,dx_S,
\end{aligned}
\label{ffpf:channel}
\end{equation}
for $m=1,\dots,N_0$, with $\eta_m\sim\mathcal N(0,N)$ and $\sigma^{2}=|A_L|^{2}D_R$. The readout is idealized as point sampling, each detector registering the image-plane intensity at the single lattice coordinate $x_{S,m}$ over an extent small against $\Delta x$: the $\sinc^{2}$ kernel is nonnegative everywhere and vanishes only at isolated lattice points. A finite-extent pixel therefore integrates the strictly positive tails of every neighboring cell---cancellations of the kind in Eq.~\eqref{ffpf:ortho} are unavailable to a nonnegative integrand---and the exact lattice diagonality $\sinc^{2}(\pi(m-n))=\delta_{mn}$ holds for point samples alone. $B_m$ is a unit-weight $\sinc^{2}$ average concentrated at the $m$-th cell but not confined to it, its tails spanning all of $S$ with polynomial decay ($(\Omega x)^{-2}$).

For an object resolved on the lattice, $B(x_S)=\Delta x\sum_n B_n\,\delta(x_S-x_{S,n})$ with cell intensities $B_n\ge0$, the average collapses by sifting to the $m$-th cell intensity exactly and the $N_0$ occupations are the resolved degrees of freedom; a general continuum object remains unresolved, its $\sinc^{2}$ kernels overlapping between the nodes, and the decoupling holds only within the band-limited lattice idealization, becoming approximate on $S$.
The residual inter-cell overlap is the irreducible far-field blur, fixed by the propagating band $\Omega\le k_0$. The $\sinc^{2}$ response is the tightest that band admits, narrowing to a point only as $\Omega\to\infty$, where $(\Omega/\pi)\sinc^{2}(\Omega x)\to\delta(x)$, outside the propagating far field. Incoherence enters through the temporal average alone, collapsing $\langle|\G^{\star}_L j|^{2}\rangle$ to the linear Hadamard map $\F B$; Eq.~\eqref{ffpf:channel} is then the bank of scalar channels of Eq.~\eqref{ffpf:rateiso}, each cell carrying its Gaussian intensity $B_m$ about a positive mean with nonnegativity held by the $\kappa$-margin of the main text. Because the spectrum is flat, the per-cell signal-to-noise ratios are equal and the uniform allocation of input variance is optimal, so the isotropic input with the focusing front end is MI-optimal at every SNR. Under a correlated prior the conjugated covariance $V_{\F}^{\dagger}\vb{Q}_BV_{\F}$ of Eq.~\eqref{ffpf:rate} departs from a multiple of the identity, the mutual information depends nontrivially on the alignment of the singular bases, and focusing is not necessarily optimal; the correlated case is treated in the main text.

\subsection{Correlated source covariance} \label{asec:correlated_covariance}
The correlated intensity covariance of Fig.~\ref{fig:iso_vs_corr} correlates the sources in pairs,
\begin{equation*}
\vb{Q}_\rho=\begin{pmatrix}1&\rho&0&0\\ \rho&1&0&0\\ 0&0&1&\rho\\ 0&0&\rho&1\end{pmatrix},\qquad \rho=0.95,
\end{equation*}
with eigenvalues \(1+\rho,\,1+\rho,\,1-\rho,\,1-\rho\), all strictly positive: sources one and two, and sources three and four, fluctuate together with correlation \(\rho\), the two pairs mutually uncorrelated, and the eigenvectors are the sums and differences within each pair. The matrix is entrywise nonnegative and full rank, with \(\operatorname{Tr}[\vb{Q}_\rho]=4=\operatorname{Tr}[\Id]\). Thus the correlated and isotropic ensembles are power-matched. The condition number \((1+\rho)/(1-\rho)=39\) concentrates the ensemble on the two pair-sum eigenmodes. The value \(\rho=0.95\) places the ensemble in the strongly correlated regime, where the incentive to align the measurement with these pair-sum eigenmodes is largest; the dependence of the optimized structure on the correlation strength is not examined here.

\bibliography{refs}

\end{document}